\documentclass[prb,showpacs,preprint]{revtex4}
\usepackage{amsmath}
\usepackage{graphicx}
\begin{document}
\title{Effects of bonding type and interface geometry
on coherent transport through the single-molecule magnet Mn$_{12}$}
\author{Kyungwha Park$^1$, Salvador Barraza-Lopez$^{2}$,
V\'{\i}ctor M. Garc\'{\i}a-Su\'arez$^{3,4}$, and Jaime Ferrer$^3$}
\affiliation{$^1$Department of Physics, Virginia Polytechnic Institute and State
University, Blacksburg VA, 24061 \\
$^2$School of Physics, Georgia Institute of
Technology, Atlanta, GA 30332 \\
$^3$Departamento de F\'{\i}sica \& CINN, Universidad de Oviedo, 33007 Oviedo, Spain \\
$^4$Department of Physics, Lancaster University, Lancaster, LA1 4YB, United Kingdom}

\begin{abstract}
We examine theoretically coherent electron transport through the single-molecule magnet Mn$_{12}$,
bridged between Au(111) electrodes, using the non-equilibrium Green's function method and the
density-functional theory. We analyze the effects of bonding type, molecular orientation, and
geometry relaxation on the electronic properties and charge and spin
transport across the single-molecule junction. We consider nine interface geometries leading to
five bonding mechanisms and two molecular orientations: (i) Au-C bonding, (ii) Au-Au bonding,
(iii) Au-S bonding, (iv) Au-H bonding, and (v) physisorption via van der Waals forces. The two
molecular orientations of Mn$_{12}$ correspond to the magnetic easy axis of the molecule
aligned perpendicular [hereafter denoted as orientation (1)] or parallel [orientation (2)] to the direction
of electron transport. We find that the electron transport is carried by the lowest unoccupied
molecular orbital (LUMO) level in all the cases that we have simulated. Relaxation of the junction
geometries mainly shifts the relevant occupied molecular levels toward the Fermi
energy as well as slightly reduces the broadening of the LUMO level. As a result, the current
slightly decreases at low bias voltage. Our calculations also show that placing the molecule in the
orientation (1) broadens the LUMO level much more than in the orientation (2), due to the internal
structure of the Mn$_{12}$. Consequently, junctions with the former orientation yield a higher current
than those with the latter. Among all of the bonding types considered, the Au-C bonding gives rise
to the highest current (about one order of magnitude higher than the Au-S bonding), for a given 
distance between the electrodes. The current through the
junction with other bonding types decreases in the order of Au-Au, Au-S, and Au-H. Importantly,
the spin-filtering effect in all the nine geometries stays robust and their ratios of the majority-spin
to the minority-spin transmission coefficients are in the range of 10$^3$ to 10$^8$. The general
trend in transport among the different bonding types and molecular orientations obtained from this
study may be applied to other single-molecular magnets.

\end{abstract}
\date{\today}
\maketitle


\section{Introduction}


Advances in experimental techniques have led to a great number of experimental studies on electron transport
through molecular junctions formed by single molecules bridged between electrodes or molecular
monolayers adsorbed onto surfaces, using three-terminal set-ups or scanning tunneling microscopy (STM) measurements.
Recently, molecular junctions based on single-molecule magnets (SMMs) connected to electrodes or monolayers of SMMs
at surfaces, have been fabricated and their electron transport characteristics \cite{HEER06,JO06,HEND07,BURG07,VOSS08,BOGA09}
have been measured, as well as their mechanical, electronic, and magnetic properties \cite{SALM07,ZOBB05,DELP06,VOSS07,OTER09,MANN09,MANN09-2}.
Electron transport through an SMM drew a lot of attention because of the intriguing interplay between its transport
properties and the internal magnetic degrees of freedom, which is absent in transport through small organic molecules.
An SMM consists of several transition metal ions interacting through organic or inorganic ligands via super-exchange
interactions. The spin configuration of the ground state of an SMM is determined by a delicate balance among the
super-exchange interactions of different strengths between the transition metal ions. Thus, the magnetic structure
of an SMM must be taken into account in understanding its electron transport and other quantum properties. Recently,
first-principles calculations of transport through an SMM were performed on a prototype SMM Mn$_{12}$ terminated
with a thiol (-S) group within Au electrodes.\cite{SALV09-1,SALV09-2,PEMM09} The calculations carried out in
Refs.[\onlinecite{SALV09-1,SALV09-2}] suggest that the Mn$_{12}$ molecule can function as a spin filter with low
bias voltage even with non-magnetic electrodes. Even though the Mn$_{12}$ is chemically bonded to the Au electrodes,
the broadening of the relevant molecular orbitals is so small compared to its charging energy that the Kondo
temperature is expected to be extremely low. In addition, these \cite{SALV09-1,SALV09-2} and other 
calculations \cite{PEMM09,MICH08} demonstrate the importance of the
internal magnetic degrees of freedom of the Mn$_{12}$ in electron transport, in contrast to typical quantum dots,
regardless of the specific details of the coupling of the molecule to the electrodes.


Systematic studies on molecular junctions based on small organic molecules reveal that the current through the
molecular junctions is highly sensitive to properties of interfaces between the molecules and electrodes because
the interfaces determine the degree of the overlap between the molecular levels and conduction channels of
the electrodes.\cite{LIAN02,PARK02,XU03,BASC05,VENK06,VENK06-2,LI06}
A thiol group is most commonly used to build a strong chemical link between the single molecules and the
Au electrodes in molecular junctions. The conductance of such single-molecule junctions is typically several
orders of magnitude smaller than $G_0=2 e^2/h$ (the conductance quantum). Binding of the S atoms
to the hollow sites of the Au surface gives rise to different conductance (by a factor of about 2 to 3) from
binding of the same S atoms to on-top sites of the Au surface.\cite{BASC05} Molecular junctions terminated with
an amine (-N) group within Au electrodes reveal even lower conductance than those with a thiol group.\cite{VENK06}
In order to increase conductance of molecular junctions, linker molecules including Au atoms \cite{MILL07} or
isocyanide derivatives \cite{BEEB02} were used between the single molecules and electrodes. In addition,
recent experiments show that conductance through
molecular junctions based on small single molecules can be enhanced to an order of $G_0$ by their direct bonding to
the electrodes without a thiol group or any other linker molecules.\cite{KIGU08,TAL09} This enhancement is
attributed to a strong coupling between the molecules and the electrodes, which places the transport in the
transparent regime rather than in the tunneling regime.\cite{TAL09,FERR09} However, corresponding systematic
studies have not yet been carried out for electron transport through an SMM. In the case of
quantum dots, the properties of interfaces are negligible in transport. However, for molecular junctions based
on SMMs, the molecules are chemically bonded to the electrodes, and thus their transport properties can change
with different interfaces even if the molecules are weakly coupled to the electrodes in the sense that the
charging energy is much greater than the broadening of the relevant molecular levels. In this paper, we investigate
how bonding types, interface geometries, and geometry relaxation influence transport characteristics of an SMM
Mn$_{12}$ when it is bridged between Au electrodes.

\begin{figure}
\includegraphics[width=7.cm, height=6.4cm]{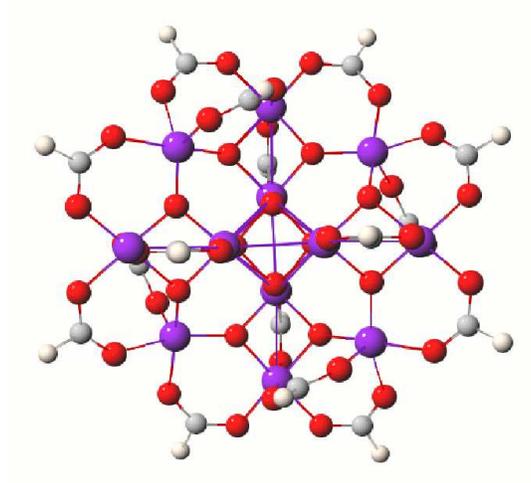}
\caption{(Color online) Top view of an isolated Mn$_{12}$ molecule: Mn (purple), O (red), C (gray),
H (white). The ground-state spin is $S=10$ or the magnetic moment is 20 $\mu_B$.}
\label{fig:isolated-Mn12geo}
\end{figure}

An SMM Mn$_{12}$ \cite{LIS80} consists of four inner Mn$^{4+}$ ions ($S=3/2$) surrounded by eight outer
Mn$^{3+}$ ions ($S=2$) which are all antiferromagnetically coupled through O anions, as well as C and H
atoms, as shown in Fig.~\ref{fig:isolated-Mn12geo}. The dominant exchange interactions are between the inner Mn$^{4+}$
ions and the outer Mn$^{3+}$ ions.\cite{PARK04} In the ground state the magnetic moments of the four Mn$^{4+}$ ions are
antiparallel to those of the eight Mn$^{3+}$ ions, such that the total spin becomes $S= 2 \times 8 - 3/2 \times 4 = 10$.
\cite{FRIE96,BARR97,HILL98}
An SMM Mn$_{12}$ initially synthesized by Lis \cite{LIS80} consists of ligands which do not form direct chemical
bonding to an Au surface. Thus, most transport experiments through an Mn$_{12}$ within Au electrodes were performed
on systems where the Mn$_{12}$ molecules are chemically bonded to surfaces or electrodes via linker molecules such as
a thiol group,\cite{HEER06,HEND07,VOSS08} or physically adsorbed onto surfaces or electrodes without any
linker molecules.\cite{JO06} To form direct Au-S bonding, some of the
original ligands of the Mn$_{12}$ molecules \cite{LIS80} were substituted by S-containing ligands, or the Mn$_{12}$
molecules were deposited onto an Au surface that was initially functionalized with S-containing alkane
chains.\cite{VOSS08} So far, the measured electric current through an Mn$_{12}$ molecule or its derivative in various
experiments is in the range of 1 pA to 100 pA at bias voltage of a few tens of mV.\cite{HEER06,JO06,VOSS08}
Compared to binding a small organic molecule to Au electrodes through a thiol or amine group, binding an Mn$_{12}$
to Au electrodes via linker molecules bears the following differences: (a) an Mn$_{12}$ has much more
binding sites to the linker molecules, and (b) the orientation of an Mn$_{12}$ relative to the electrodes is,
to a great extent, determined by binding sites of the linker molecules to the Mn$_{12}$, rather than their binding
sites to the electrodes.

Full control of the properties of interfaces is extremely difficult to achieve in molecular junctions and tunneling
measurements using STM or scanning tunneling spectroscopy (STS).\cite{XU03,VENK06,VENK06-2,LI06,KIGU08,TAL09}
Thus, in transport experiments, for a given bonding type, several binding sites are plausible, and hundreds of
fabricated samples of the bonding type give rise to a histogram of conductance for given gate voltage or a
series of current-voltage curves. In the present study, we take into account five different bonding types between
an Mn$_{12}$ and Au electrodes, and for each bonding type, some representative interface geometries are examined.
Even though we do not simulate all possible configurations of the interface that may be realized in fabricated samples
for a given bonding type, the general trend in our calculated transport properties will elucidate the effects of linker
group, molecular orientation, and geometry relaxation, as well as the effectiveness of different bonding
types and interface geometries in transport through an Mn$_{12}$ and other SMMs.

\begin{figure}
\includegraphics[width=12.cm, height=18.5cm]{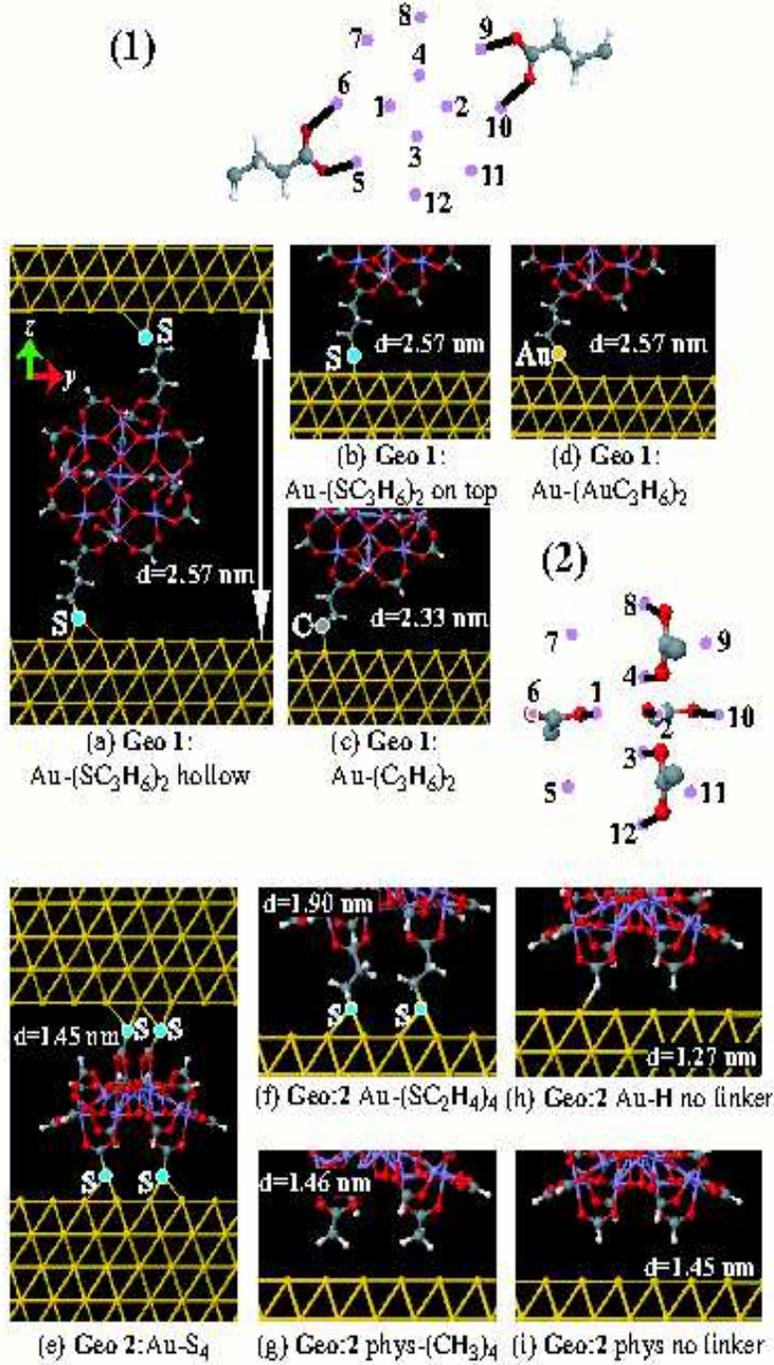}
\caption{(Color online) The nine geometries of the scattering region with different bonding types
and linker groups. Not all of the Au atomic layers are shown.
(a), (b), (e), (f): Au-S bonding with different linker molecules and in the
orientations (1) and (2). (c): Au-C bonding, (d): Au-Au bonding, (g) and (i):
physisorption with and without linker molecules, respectively. (h): Au-H bonding.
On the top and in the middle, the positions of the twelve Mn ions in relation to the
binding sites for the linker molecules are labeled for the orientations (1) and (2),
respectively.}
\label{fig:geo}
\end{figure}

In this work, we consider an Mn$_{12}$ molecule bridged between Au(111) electrodes (two-terminal set-up) via
nine different ways (Fig.~\ref{fig:geo}), and investigate their current-voltage characteristics using the
density-functional theory (DFT) and the nonequilibrium Green's function method (NEGF). The nine different ways can be
categorized according to their bonding types: (i) Au-S bonding, (ii) Au-C bonding, (iii) Au-Au bonding,
(iv) Au-H bonding, and (v) physisorption through van der Waals forces. DFT does not fully capture
the nature of van der Waals forces where the orbitals considered do not overlap. We note, however, that
if molecular orbitals do not overlap with the orbitals of electrodes at a molecular junction, the transport
properties of the junction belong to the tunneling regime. A recent study reveals that first-principles
calculations based on DFT and NEGF provide qualitative features of transport in such a regime.\cite{GARC05}
For example, when a Pt break junction becomes broken with vacuum between the electrodes, a first-principles
calculation showed exponentially decaying conductance as a function of distance between the electrodes.
\cite{GARC05} This result suggests that our methodology can be used to understand the transport in the
case of physisorption, although the physical mechanism in our case differs from that in the broken junction.
In the present calculations, we consider two molecular orientations relative to the electrodes:
when the magnetic easy axis of the Mn$_{12}$ is normal [orientation (1)] or parallel
[orientation (2)] to the transport direction, the $z$ axis, as shown in Fig.~\ref{fig:geo}. The structure
of the Mn$_{12}$ dictates that in the orientation (1) one linker molecule can be attached to each electrode
without significantly deforming the Mn$_{12}$, while in the orientation (2), two linker molecules are
possibly accommodated for each electrode without much distortion. As illustrated in Fig.~\ref{fig:geo},
the nine different interface geometries are named based on the following rule. Adapted from the notations
used in Refs.[\onlinecite{SALV09-1,SALV09-2}], we refer to the orientation (1) as Geo 1, and to the
orientation (2) as Geo 2. For each interface geometry, the orientation of an Mn$_{12}$ is first specified,
then the type of adsorption (either chemical or physical) is given. For chemical bonding, the type of bonding
and the name of the linker group are noted, while for physisorption, whether linker molecules are placed or
not is stated. For example, Geo 1:Au-(SC$_3$H$_6$)$_2$-hollow [Fig.~\ref{fig:geo}(a)] denotes an interface
geometry with the Mn$_{12}$ in the orientation (1) and the linker molecule (SC$_3$H$_6$)$_2$ bonded to the
hollow sites of the electrodes via Au-S bonding. We will first outline our model and
computational method. Then we will discuss the interface geometries in detail, and the effects of
geometry relaxation, molecular orientation, bonding type, and linker group on transport properties
of the Mn$_{12}$. A brief conclusion will follow.



\section{Computational method and model}

We use the quantum transport code {\tt SMEAGOL} \cite{SMEAGOL,FERN06} interfaced with the DFT code
{\tt SIESTA}\cite{SIESTA}. The Au electrodes are treated semi-infinitely
using {\tt SIESTA}. The scattering region consists of the Mn$_{12}$, linker molecules, and six flat
Au atomic layers on each side of the Mn$_{12}$ ($6 \times 6$ surface atoms per layer), as shown in
Fig.~\ref{fig:geo}. Even if the number of the Au atomic layers increases from six to nine on each
side of the Mn$_{12}$, the transmission probability does not change at all. This was tested and
confirmed for the interface geometry with the Au-C bonding [Geo 1:Au-(C$_3$H$_6$)$_2$, Fig.~\ref{fig:geo}(c)],
where the broadening of the relevant molecular level turns out to be the largest
among the interface geometries considered. The electronic structures of the electrodes and scattering
region are calculated within the spin-polarized Perdew-Burke-Ernzerhopf (PBE) generalized-gradient approximation
(GGA) \cite{PERD96}, using {\tt SIESTA}. We generate Troullier-Martins pseudopotentials \cite{TROU91} for Au, Mn, S, O,
C, and H with scalar relativistic terms and core corrections except for H. We also build corresponding basis
sets of the elements using the scheme presented in Ref.~[\onlinecite{JUNQ01}]. For the Mn basis set,
3$p$ orbitals as a semicore are required in order for the molecular orbital levels of an isolated Mn$_{12}$
to be comparable to those obtained from all-electron calculations \cite{PEDE99}. With these pseudopotentials
and basis sets, self-consistent DFT calculations including spin-orbit coupling \cite{FERN06} are performed
for an isolated Mn$_{12}$. Using a modified version \cite{FERN06} of {\tt SIESTA}, 
we obtain the total magnetic moment of 20 $\mu_B$ and the 
magnetic anisotropy barrier of 66.4~K,\cite{SALV09-1}. This is in good agreement with experiment \cite{BARR97} 
and with the barrier \cite{SALV07,SALV08} computed using the DFT code {\tt VASP}\cite{VASP}. 
To reduce the computational cost, a small Au basis set of a single $s$ orbital is used for transport 
calculations, while a large Au basis set of both $d$ and $s$ orbitals is used for geometry relaxation.
The transmission at the Fermi level in Au is dominated by the $s$ states because 
the $d$ orbital levels are well below the Fermi level.
It is checked that in the scattering region, the density of states projected onto Mn $d$ orbitals with
the large Au basis set is fairly similar to that with the small Au basis set, which justifies the
utilization of the small Au basis set in the transport study. For a given interface geometry, the 
distance $d$ between the electrodes was determined such that all of the terminating atoms from the linker molecules (such as S, C, 
or Au atoms) are bonded to the lowest-energy sites of the the Au slab with reasonable bond lengths. 
For example, for the Au-S bonding, the distance $d$ was obtained from an optimum bond length between a 
S-terminated alkane chain and the hollow site of a small Au cluster. In constructing the initial geometry 
of the scattering region, each component of the region, such as the electrodes, the Mn$_{12}$, and the 
linker molecules, are separately optimized in advance. The largest force is found at the interface between
the linker molecules and the electrodes.\cite{SALV07} Thus, the geometry relaxation with a fixed $d$ allows
the linker molecules and the Mn$_{12}$ to relax further and lowers the force at the interface. 
Prior to the transport calculations, all of 
the geometries of the scattering region or interface geometries considered are relaxed with a fixed 
distance $d$ between the electrodes, using {\tt SIESTA}, until the magnitude of the maximum force exerted on 
the atoms becomes less than 0.1 eV/\AA,~unless stated otherwise.

To avoid any quantum confinement effects,\cite{SMEAGOL} for the scattering region, periodic boundary 
conditions are applied in the transverse directions and $3 \times 3 \times 1$ $k$-points are sampled. 
The retarded Green function for the scattering region, $G_{\rm{EM}}^{\rm{R}}$, has the form of
$[\epsilon^{+} S_{\rm{EM}} - H_{\rm{EM}} -\Sigma_{\rm L}^{\rm R} - \Sigma_{\rm R}^{\rm R}]^{-1}$, where
$S_{\rm{EM}}$ and $H_{\rm{EM}}$ are the overlap and Hamiltonian matrices for the scattering region,
respectively.\cite{SMEAGOL} Here $\Sigma_{\rm L}^{\rm R}$ and $\Sigma_{\rm R}^{\rm R}$ are self-energies 
arising from the interactions of the Mn$_{12}$ with the left and right electrodes, respectively. 
These self-energies depend on energy $E$ and $k$ vector, and they are obtained from spin-polarized 
calculations of retarded surface Green functions for the electrodes which are computed using
the scheme developed in Ref.[\onlinecite{SANV99}]. The initial magnetic moment of the scattering region 
is set to 20 $\mu_B$. Interactions with phonons or additional 
electron correlations such as on-site Coulomb repulsion (Hubbard-like $U$ term) are not taken into 
account. The density matrix of the scattering region is self-consistently solved within the NEGF formalism 
until it converges. After the convergence of the density matrix, the total magnetic moment slightly 
increases, such as 20.3 $\mu_B$ for Geo 1:Au-(SC$_3$H$_6$)$_2$-hollow, due to a small amount of spin 
polarization in the Au atomic layers caused by the Mn$_{12}$.
Then the transmission $T(E, V_b)$ at low bias voltage $V_b$ is calculated 
as follows:
\begin{eqnarray}
T(E,V_b)&=& {\rm Tr}[\Gamma_L \: G^{{\rm R} \dag}_{\rm EM} \: \Gamma_R \: G^{\rm R}_{\rm EM}](E,V_b),
\end{eqnarray}
where $\Gamma_L$ and $\Gamma_R$ denote the broadening of molecular levels induced by coupling to the left and
right electrodes. Accurate calculations of $T(E, V_b)$ require high resolution in energy $E$ (such as
$7 \mu$eV for very sharp transmission peaks) due to the weak coupling between the Mn$_{12}$ and the electrodes. Since
the electrodes are treated semi-infinitely and periodic boundary conditions are employed, the 
transmission coefficients are obtained after their integration over the $k$-points.
The current $I$ as a function of $V_b$ is obtained from
\begin{eqnarray}
I (V_b) &=& \frac{e}{h} \: \int dE \: \: T(E,V_b) \: [f(E+eV_b/2) - f(E-eV_b/2)],
\label{eq:IV}
\end{eqnarray}
where $f(E+eV_b/2)$ and $f(E-eV_b/2)$ are the Fermi-Dirac distribution functions of the left and right
electrodes. The electronic temperature used is 10 K.

\section{Results and Discussion}


We examine the nine interface geometries (Fig.~\ref{fig:geo}) leading to the five bonding types and two
molecular orientations that were discussed in Sec.~I. Their molecular orientations, linker groups,
bonding types, binding sites, and distances $d$ between the electrodes are listed in Table~1.

\subsection{Interface geometries considered}

In the Au-S bonding, four interface geometries are considered: Fig.~\ref{fig:geo}(a), (b), (e), and (f).
In Fig.~\ref{fig:geo}(a), Geo 1:Au-(SC$_3$H$_6$)$_2$-hollow, one SC$_3$H$_6$ linker molecule is bonded to
each Au electrode, where the S atom is adsorbed at the hollow sites of the flat Au surface. For a fixed
distance $d$, after the geometry relaxation, this geometry bears the shortest distances between the S and Au
surface atoms in the range of 2.52 to 2.82~\AA.~If the S atom is bonded to the on-top site of the Au surface
[Geo 1:Au-(SC$_3$H$_6$)$_2$-ontop, Fig.~\ref{fig:geo}(b)], the total energy increases by 0.094~eV
compared to the hollow-site case, and the shortest distance between the S and Au atoms becomes 1.92~\AA.
In Fig.~\ref{fig:geo}(e) and (f), two S linker atoms [Geo 2:Au-S$_4$] or two SC$_2$H$_4$ linker molecules
[Geo 2:Au-(SC$_2$H$_4$)$_4$] are bonded to the hollow sites of the surface of each electrode. For these
two geometries the shortest distances between the S and Au atoms are in the range of 2.52 to 2.68~\AA.

Our study of the Au-C bonding was motivated by recent measurements on the conductance of a molecular
junction based on benzene directly bonded to Pt electrodes.\cite{KIGU08} It is prevalent that chemical
bonding between a metallic surface and C atoms is difficult to achieve under ordinary conditions, especially
between an Au surface and C atoms. However, the direct Au-C bonding does occur in a few
exceptional cases, such as isocyanide derivatives adsorbed on Au surfaces.\cite{GRUE06,DUBO06}
In the geometry with the Au-C bonding [Geo 1:Au-(C$_3$H$_6$)$_2$, Fig.~\ref{fig:geo}(c)], one
C$_3$H$_6$ linker molecule is adsorbed at the on-top site of each electrode, and the shortest distance
between the Au and outermost C atoms is 2.28 (2.37)~\AA~for the left (right) linker molecule.
The study of the Au-Au bonding was inspired by an experimental effort to increase conductance through
small molecules using Au-Au bonding.\cite{MILL07} In the geometry with the Au-Au bonding
[Geo 1:Au-(AuC$_3$H$_6$)$_2$, Fig.~\ref{fig:geo}(d)], one
AuC$_3$H$_6$ linker molecule is bonded to the hollow sites of each electrode, and the shortest distances
between the Au surface and the Au atoms from the linker molecules are in the range of 2.73 to 2.79~\AA.

The study of the geometries with physisorption and without linker molecules, was motivated by the transport
measurement through the Mn$_{12}$ without any linker molecules \cite{JO06}. We consider such geometries
in the orientation (2). In the geometry with the Au-H bonding [Geo 2:Au-H-no-linker, Fig.~\ref{fig:geo}(h)],
no linker molecules are placed between the Mn$_{12}$ and the electrodes. However, the short distance $d$ in that
geometry allows Au-H bonding to be initially formed at four different locations. We choose one of the four
Au-H bonding distances in its initial geometry to be slightly shorter than the three distances, and then
upon geometry relaxation, only one Au-H bonding remains effective as shown in Fig.~\ref{fig:geo}(h).
In the case of physisorption, two geometries are considered: Figs.~\ref{fig:geo}(g) and (i).
In Fig.~\ref{fig:geo}(g), Geo 2:phys-(CH$_3$)$_4$, two CH$_3$ linker molecules
are attached to each side of the Mn$_{12}$. Upon geometry relaxation (with a fixed $d$ until the magnitude of
the maximum force is less than 0.1~eV/\AA),~in this geometry, the shortest distance between the Au
and the C atoms from the linker molecules is 2.41~\AA,~which is long enough such that chemical bond is
not formed between the linker molecules and the electrodes. In Fig.~\ref{fig:geo}(i), Geo 2:phys-no-linker, no linker molecules
are placed between the Mn$_{12}$ and the electrodes, and the Au surface atoms are well separated from the H
atoms from the Mn$_{12}$ in the range of 2.97 to 3.10~\AA.~Thus, no chemical bond is formed between
the electrodes and the Mn$_{12}$ in this geometry, either. The geometry of Geo 2:phys-no-linker was built
from the optimized geometry of Geo 2:Au-H-no-linker with an increased distance $d$. Hereafter, when a particular
bonding type is discussed, its corresponding binding site is also implicitly assumed.



\subsection{Effect of geometry relaxation}

For an isolated Mn$_{12}$ molecule in the gas phase, without the on-site Coulomb repulsion $U$, our DFT
calculations using {\tt SIESTA} show that the energy gap between the highest occupied molecular orbital
(HOMO) and the lowest unoccupied molecular orbital (LUMO) levels is 0.38~eV. Since the Mn$_{12}$ has
a high magnetic moment of 20 $\mu_B$ in the ground state, both the HOMO and LUMO arise
from the majority-spin Mn $d$ orbitals. This feature of the HOMO and LUMO does not change with inclusion
of a proper value of the $U$ term, although the HOMO-LUMO energy gap increases with the $U$ term. This
was shown in previous DFT+U calculations \cite{SALV08,BOUK07}, in contrast to the result discussed in
Ref.[\onlinecite{PEMM09}]. Such a characteristic of the HOMO and LUMO was also earlier revealed by
all-electron DFT calculations\cite{PEDE99}, and it was indirectly demonstrated in experiments on
locally charged Mn$_{12}$ molecules\cite{EPPL95,BASL05}.

\begin{figure}
\includegraphics[width=7.8cm, height=6.0cm]{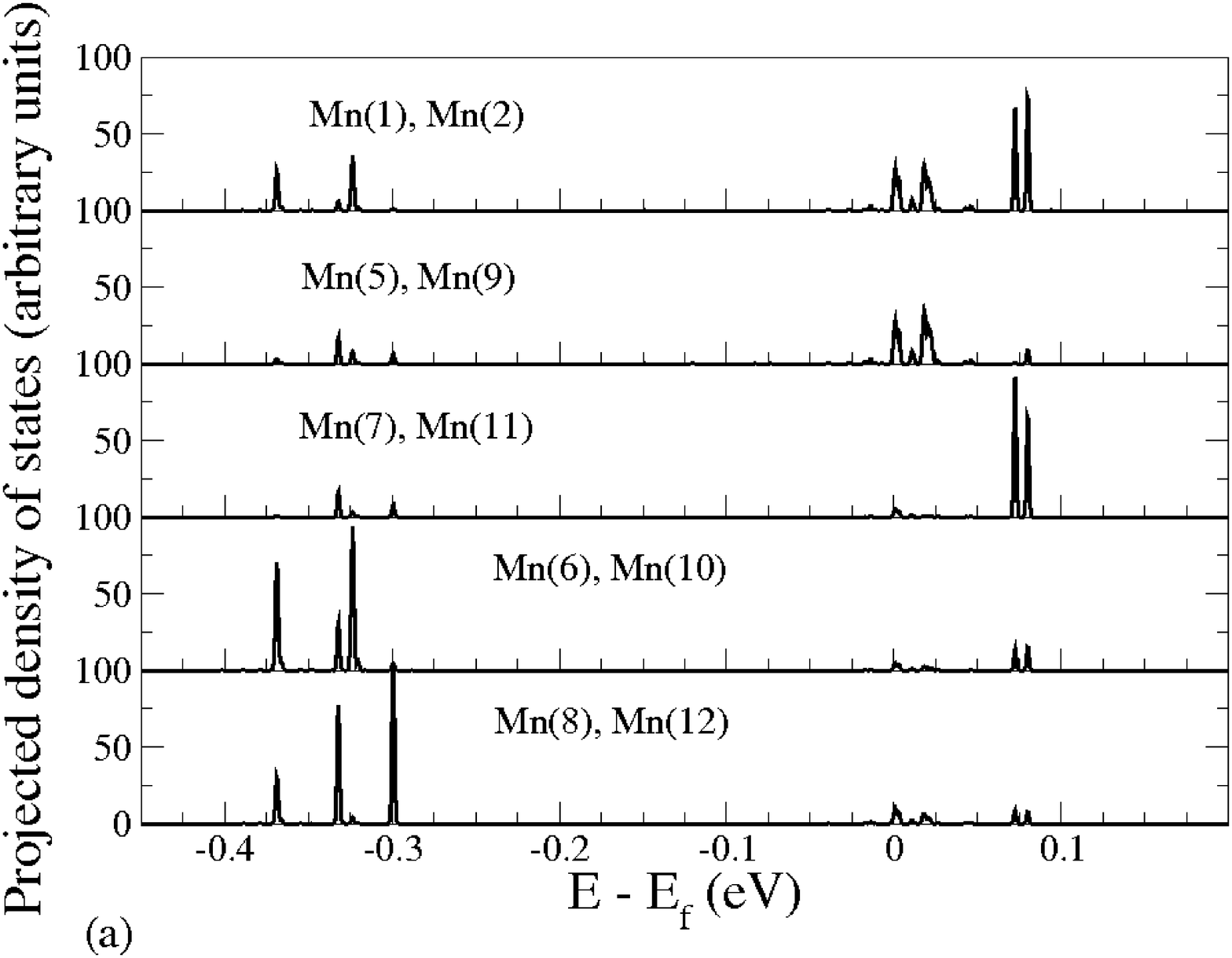}
\hspace{0.5truecm}
\includegraphics[width=7.8cm, height=6.0cm]{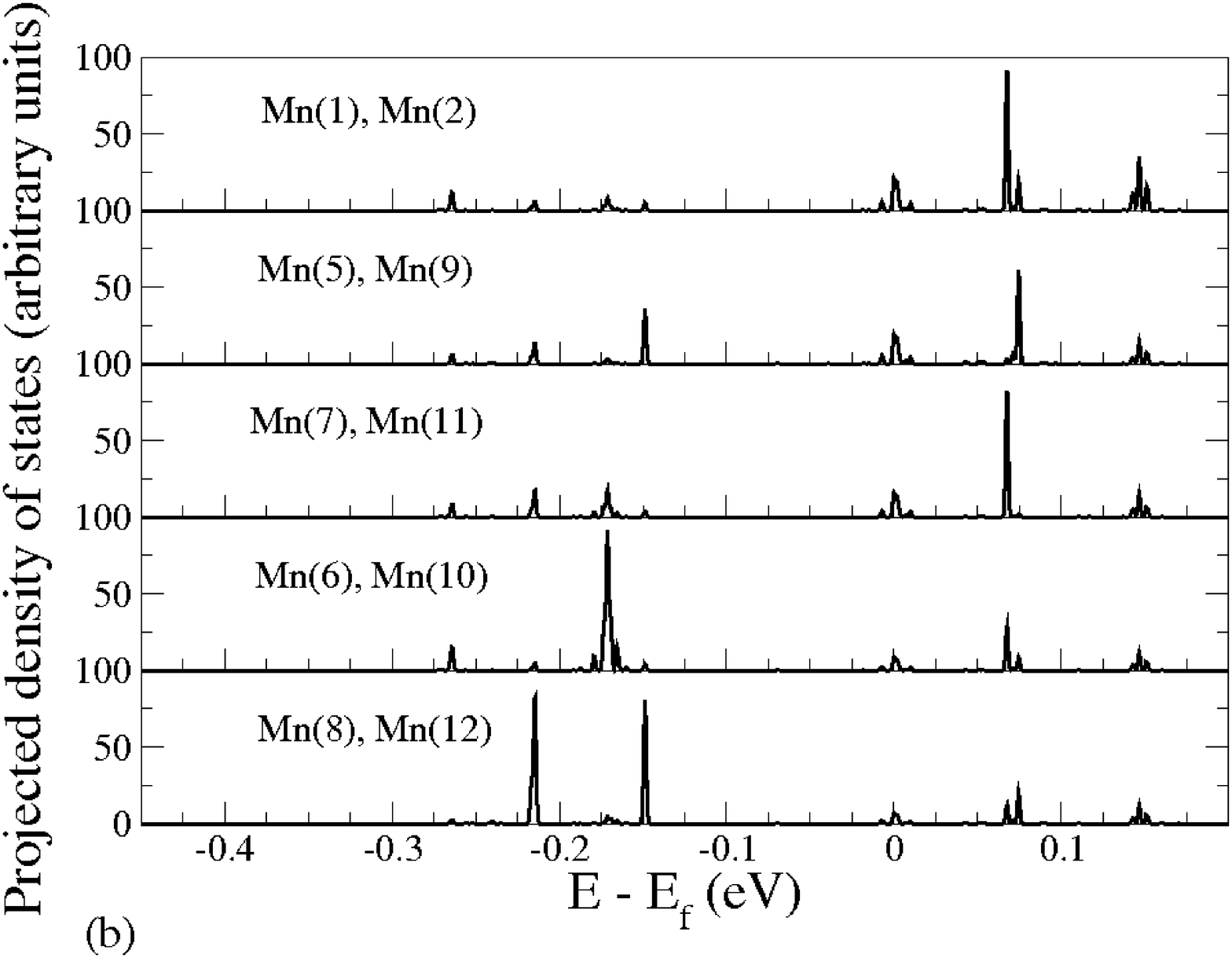}
\caption{Majority-spin density of states projected onto the Mn $d$ orbitals
of the (a) initial and (b) optimized geometries, Geo 1:Au-(SC$_3$H$_6$)$_2$-hollow.
Refer to Fig.~\ref{fig:geo} for numbering of the Mn sites. The densities
of states for Mn(3) and Mn(4) sites are the same as those for Mn(1) and Mn(2). }
\label{fig:PDOS-geo1}
\end{figure}

When an Mn$_{12}$ is bridged between the electrodes in the orientation (1), the S$_4$ symmetry of
an isolated Mn$_{12}$ is broken and the molecular levels in the interface show approximately twofold
symmetry. In addition, the magnetic easy axis of the Mn$_{12}$ is slightly tilted from the axis perpendicular
[$x$ axis, Fig.~\ref{fig:geo}(a)] to the transport direction, in order to allow the linker molecules
to be bonded to the electrodes. Furthermore, geometry relaxation at a fixed $d$ renders some noticeable
changes in the molecular levels. To compare the molecular levels before and after the geometry relaxation,
we examine the geometry Geo 1:Au-(SC$_3$H$_6$)$_2$-hollow. For its initial geometry, the HOMO and LUMO
levels of the Mn $d$ orbitals, are separated by 0.3~eV, which is similar to the HOMO-LUMO energy gap of
0.38~eV for the isolated Mn$_{12}$. With the geometry optimization,
the Mn $d$ orbital levels right below the Fermi level $E_f$ approach toward $E_f$, but the Mn $d$ levels right
above $E_f$ become split (Fig.\ref{fig:PDOS-geo1}). As a result, the separation between the HOMO and LUMO
levels reduces to 0.15~eV. The HOMO level is located near $-0.15$~eV, while the LUMO level is near $E_f$
[Fig.~\ref{fig:PDOS-geo1}(b)] (this is due to some charge transfer from the Au electrodes to 
the Mn$_{12}$ \cite{SALV07}). The same kind of shift is expected for the rest of the geometries
considered in the orientation (1). Now when an Mn$_{12}$ is bridged between the electrodes in the orientation
(2), the S$_4$ symmetry is by large preserved in the molecular levels in the interface. If the
distance $d$ is not too short, the geometry relaxation will shift the molecular levels for geometries
in the orientation (2) as well. One exception is Geo 2:Au-S$_4$, where a combination of the strong Au-S bonding
with the short distance $d$ prevents the Mn$_{12}$ and the four S atoms from being significantly rotated or
stretched/compressed with geometry relaxation.

\begin{figure}
\includegraphics[width=7.8cm, height=6.0cm]{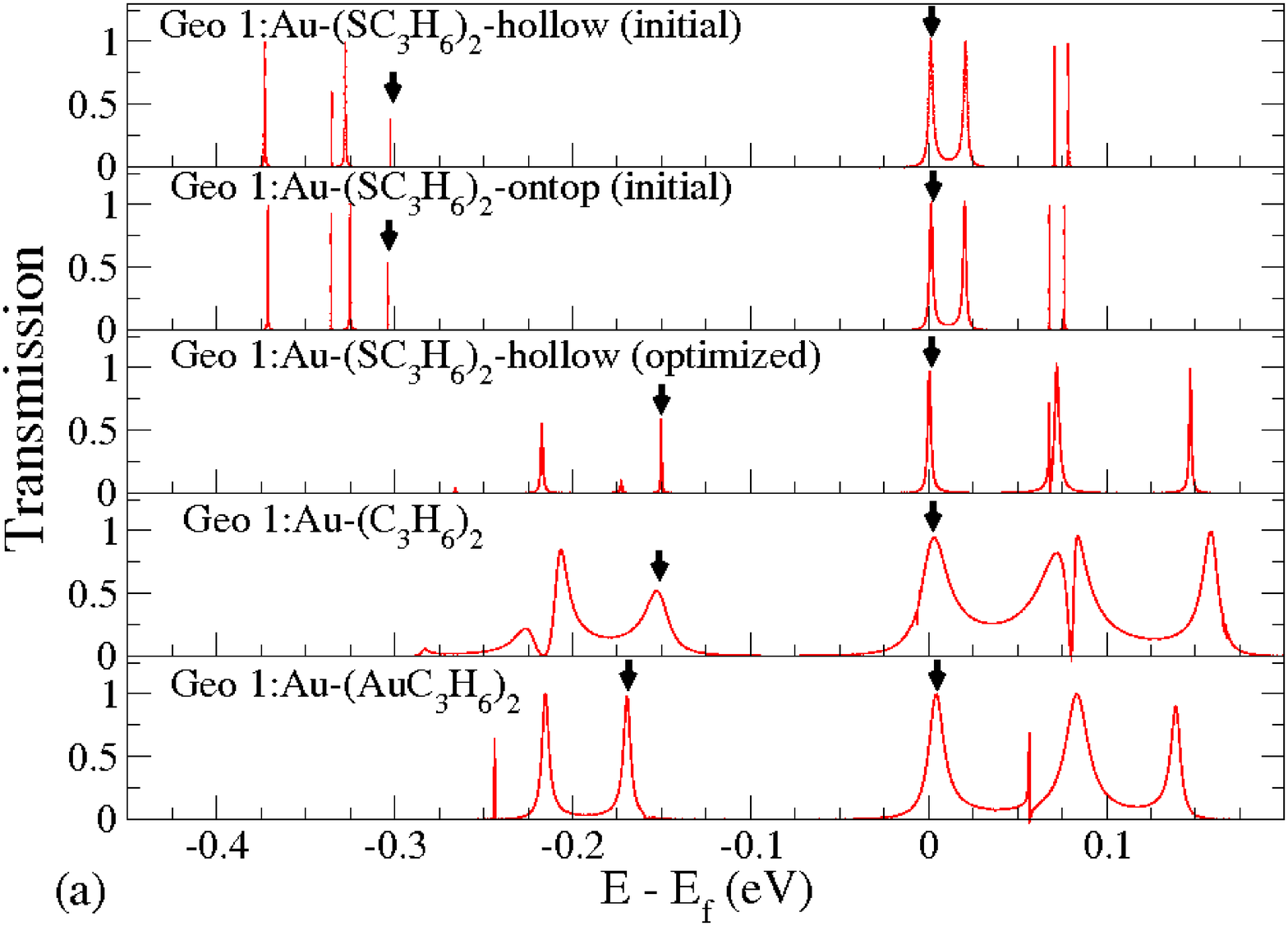}
\hspace{0.5truecm}
\includegraphics[width=7.8cm, height=6.0cm]{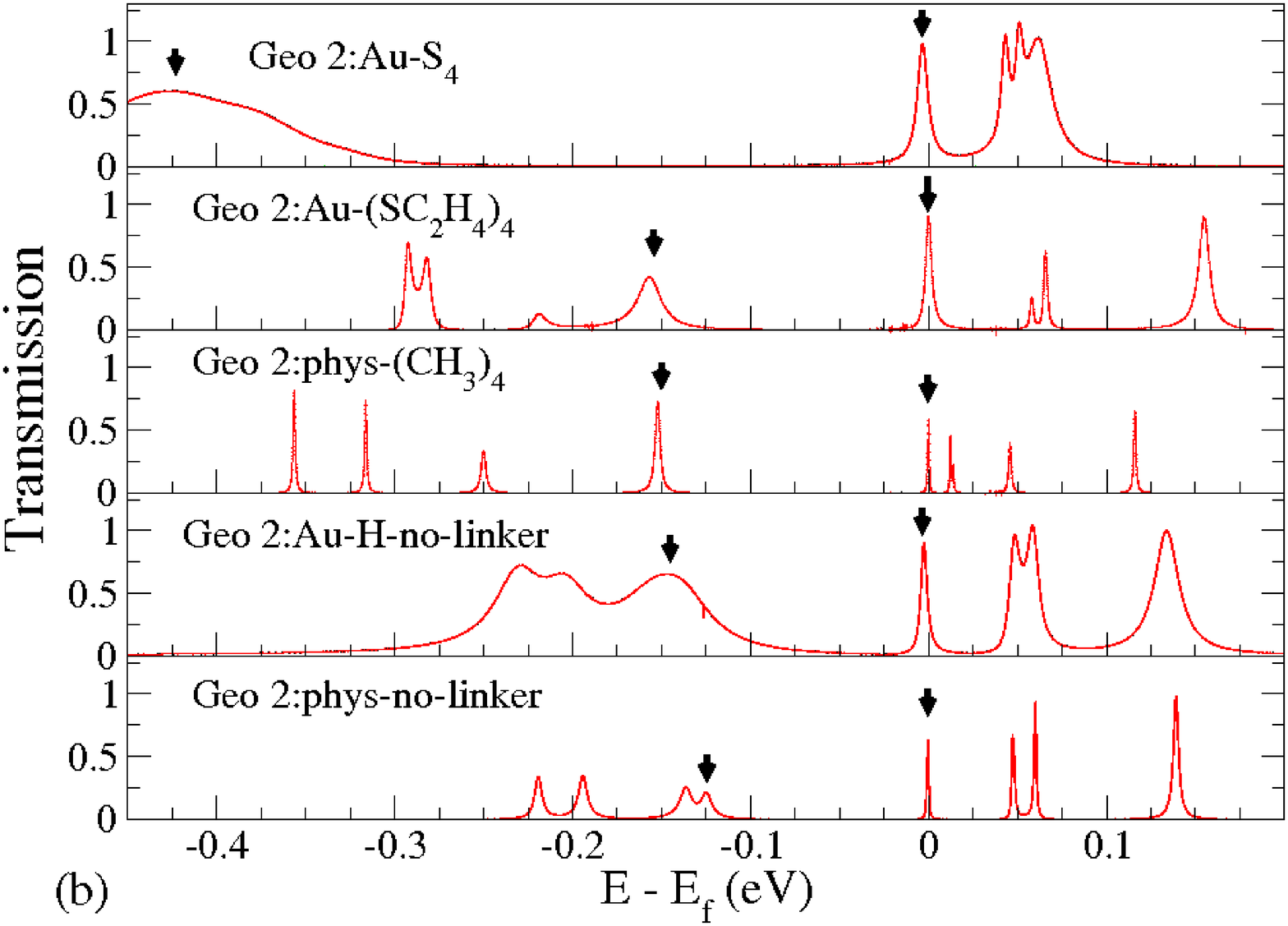}
\caption{(Color online) (a) Majority-spin transmission $T(E,0)$ at zero bias
as a function of energy $E$ relative to the Fermi level $E_f$ for the
five interface geometries with the orientation (1), and (b) for the five
geometries with the orientation (2). The arrows indicate the transmission
peaks associated with the HOMO and LUMO levels of the Mn$_{12}$.}
\label{fig:TRC}
\end{figure}

\begin{figure}
\includegraphics[width=16.cm, height=12.cm]{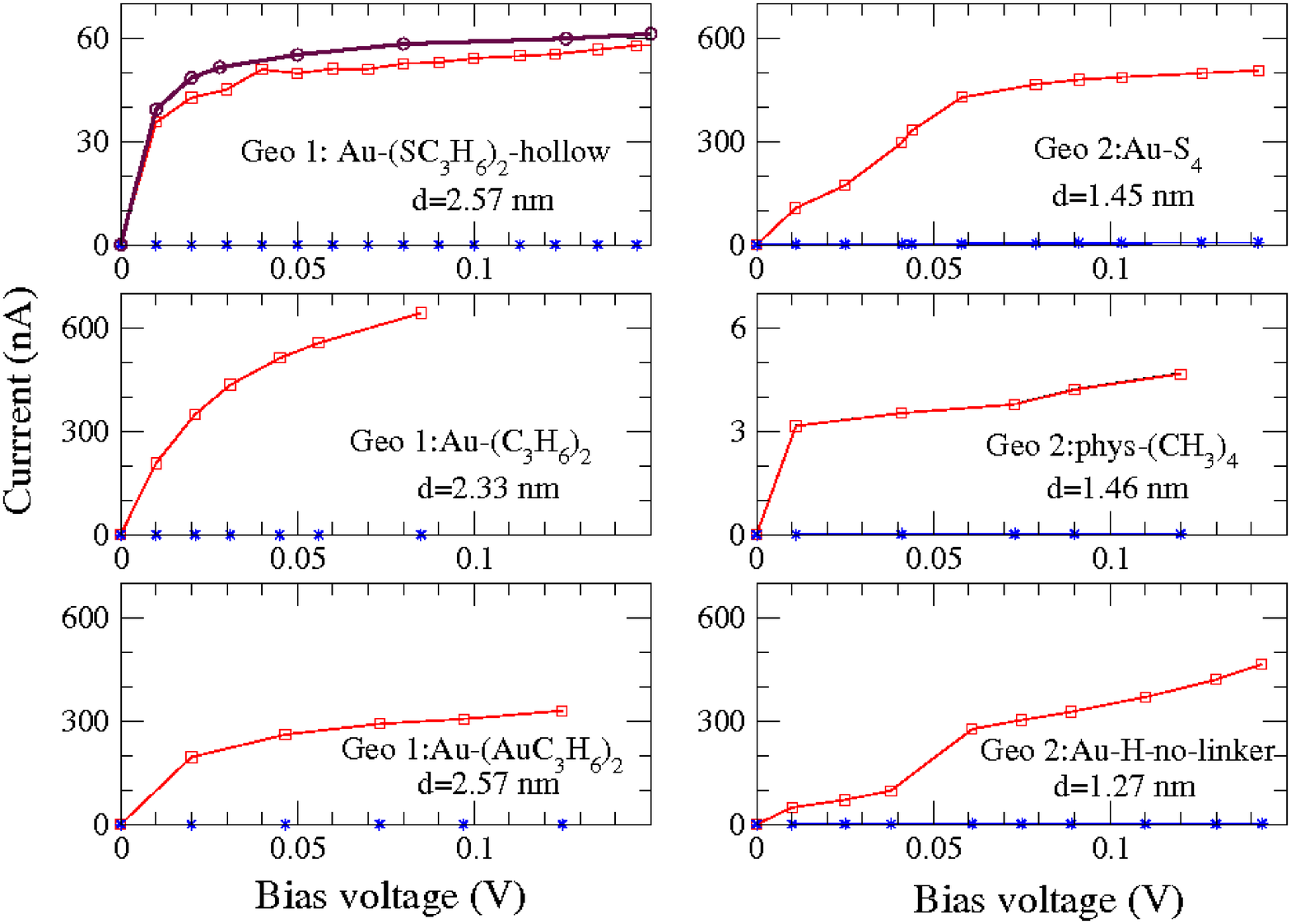}
\caption{(Color online) Current-voltage ($I-V$) characteristics for six interface geometries:
majority-spin contribution (red box), minority-spin contribution (blue star). The
topmost left panel shows $I-V$ curves for both the initial (brown circle)
and the optimized geometries Geo 1:Au-(SC$_3$H$_6$)$_2$-hollow (red box). Note that
Geo 1:Au-(SC$_3$H$_6$)$_2$-hollow and Geo 2:phys-(CH$_3$)$_4$ have different vertical scales
from the rest of the geometries.}
\label{fig:IV}
\end{figure}

We discuss the effect of geometry relaxation on $T(E,0)$. Similarly to the effect on the molecular levels,
the geometry relaxation gives rise to a upward shift in the transmission peaks right below $E_f$ and
more spread peaks right above $E_f$, as shown in the first and the third panels of Fig.~\ref{fig:TRC}(a)
from the top.
According to our calculations, the current is carried via the LUMO level of the Mn$_{12}$, and thus
the broadening of the LUMO level is relevant to the current-voltage characteristics. Without contributions
from interactions with phonons, in resonance, the individual transmission peak centered at $E_i$ for given $V_b$
(for a single conductance channel) is
solely described by the broadening of the corresponding molecular level as follows:\cite{DATT95,DATT04}
\begin{eqnarray}
T(E,V_b) &=& \frac{\Gamma_1(E_i,V_b) \Gamma_2(E_i,V_b) }{4(E-E_i)^2 + \Gamma(E_i,V_b)^2 },
\label{eq:TRC}
\end{eqnarray}
where $\Gamma_1(E_i,V_b)$ and $\Gamma_2(E_i,V_b)$ are the broadening of the $i$th molecular level
caused by the left and right electrodes, respectively, and they are determined by full widths at
half maximum of the transmission peak centered at $E_i$. Here $\Gamma(E_i,V_b)$ is the average broadening
given by $(\Gamma_1(E_i,V_b) + \Gamma_2(E_i,V_b))/2$. (Notice that for symmetric coupling,
$\Gamma_1=\Gamma_2$, Eq.~(\ref{eq:TRC}) implies that $T$ becomes unity at $E=E_i$.)
The orbital broadening $\Gamma_{1,2}(E_i,V_b)$ is
related to the hopping integral $t_{1,2}$ (between the electrodes and the Mn$_{12}$) as
$\pi t_{1,2}^2 \rho(E_i)$, where $\rho(E_i)$ is the density of
states at the $i$th level.\cite{ZIMA72} The values of $t_{1,2}$ and $\rho(E_i)$ depend on bonding types and
binding sites. We compute the values of $\Gamma_1$ and $\Gamma_2$ from fitting the
transmission peaks to Eq.~(\ref{eq:TRC}). For Geo 1:Au-(SC$_3$H$_6$)$_2$-hollow, the geometry relaxation
reduces the value of $\Gamma_{\rm{LUMO}}$ from 0.0028 to 0.0020~eV, while it enhances the value of
$\Gamma_{\rm{HOMO}}$ from 2.6$\times 10^{-5}$ to 5.0$\times 10^{-4}$~eV, as listed in Table~1. Since
the value of $\Gamma_{\rm{LUMO}}$
decreases to a small degree with the geometry relaxation, a slightly lower current flows through the relaxed
junction than via the initial one (the topmost left panel of Fig.~\ref{fig:IV}). Similar behavior
is expected for the rest of the geometries. Henceforth, we discuss optimized interface geometries only, unless
specified otherwise.

\subsection{Effect of molecular orientation}

\begin{figure}
\includegraphics[width=7.8cm, height=4.8cm]{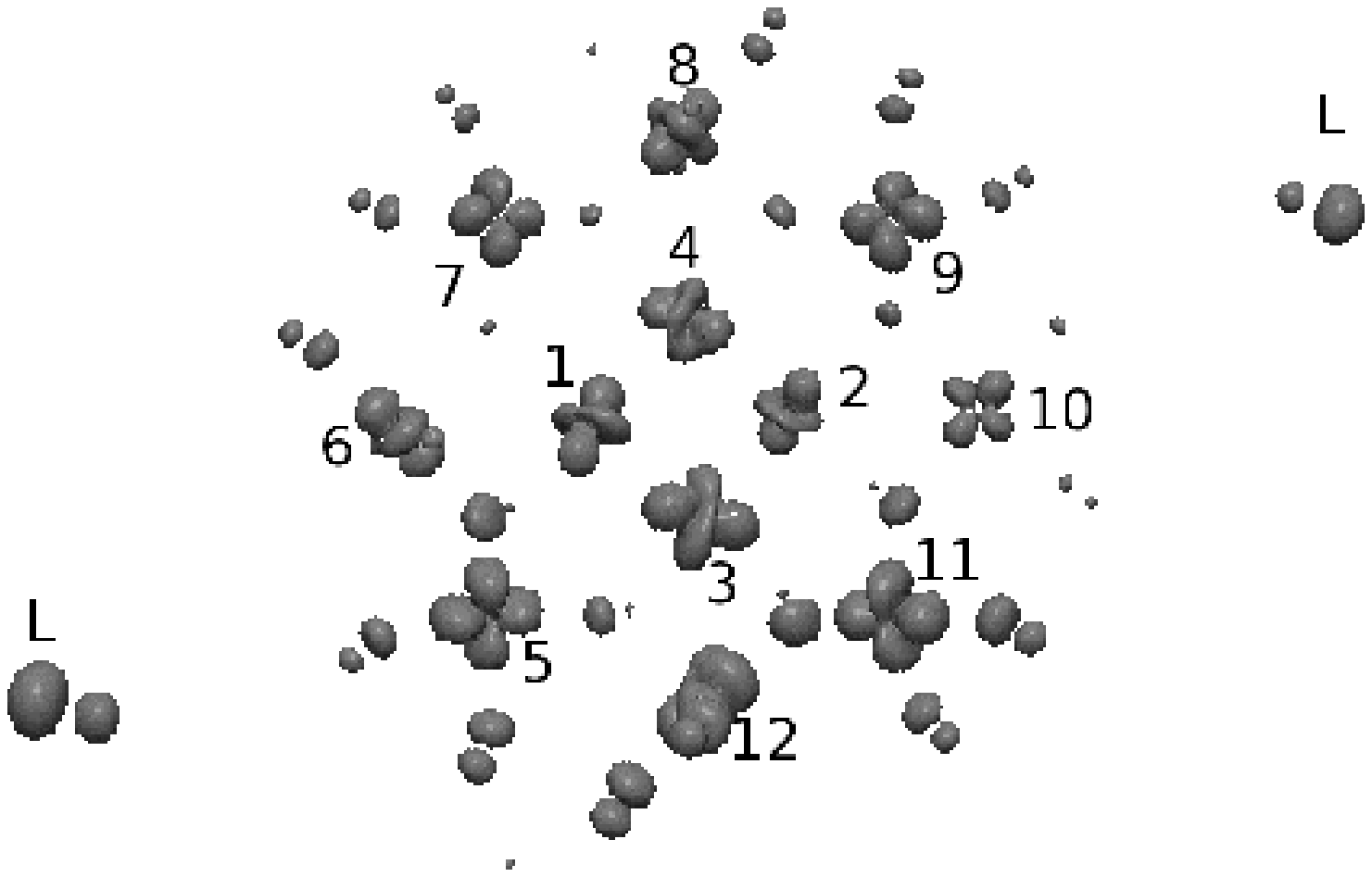}
\hspace{0.5truecm}
\includegraphics[width=7.8cm, height=4.8cm]{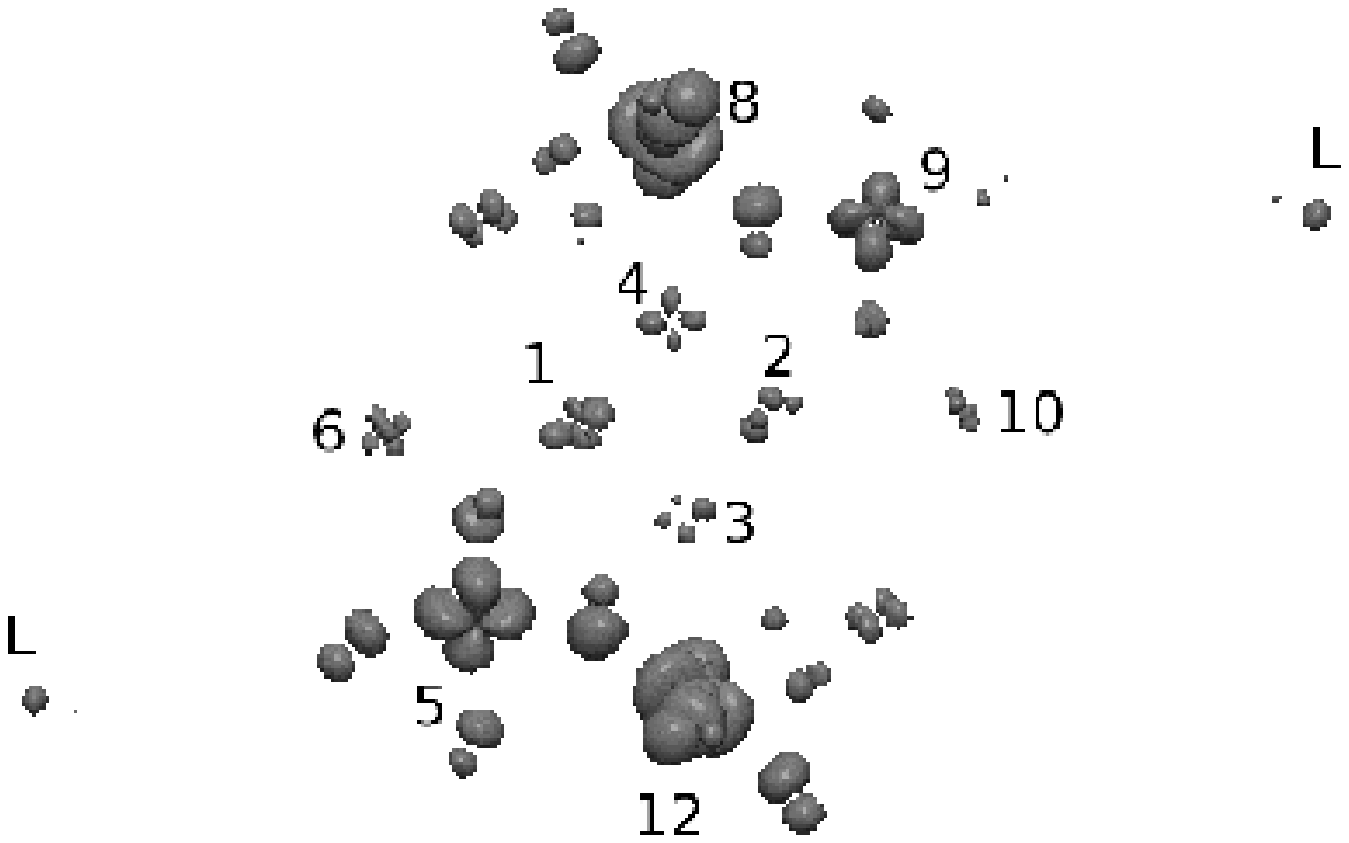}
\includegraphics[width=5.3cm, height=4.8cm]{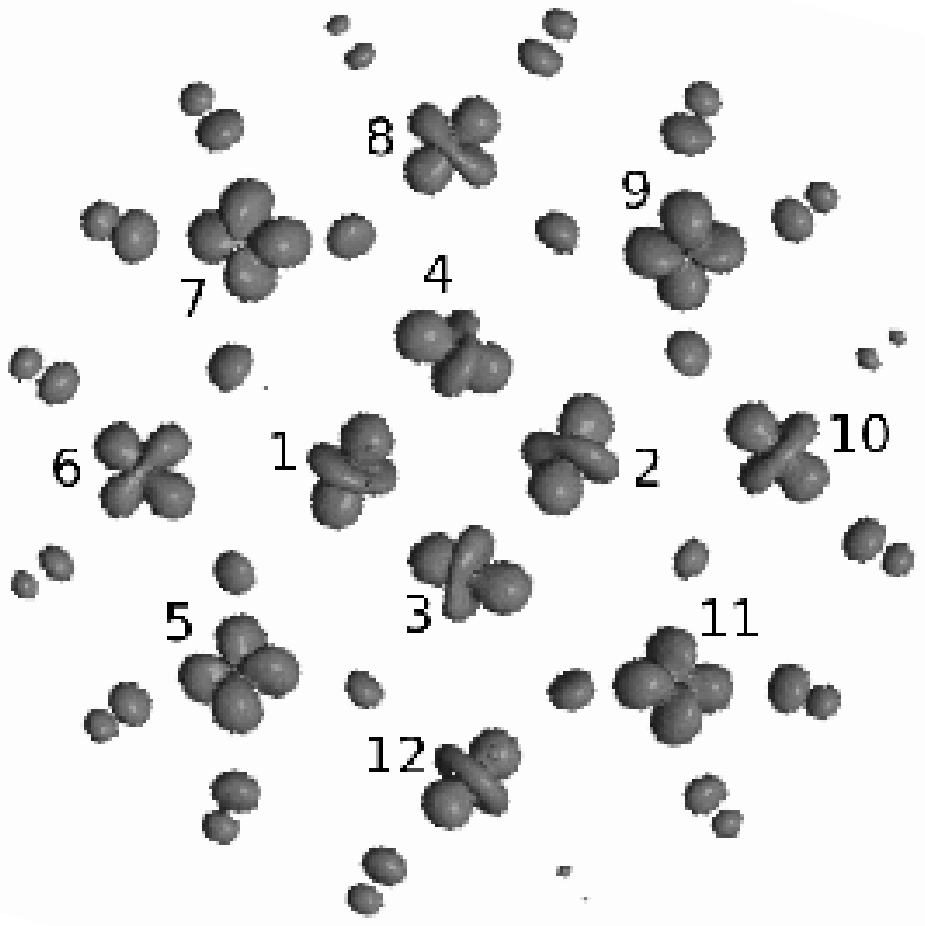}
\hspace{3.5truecm}
\includegraphics[width=4.7cm, height=4.25cm]{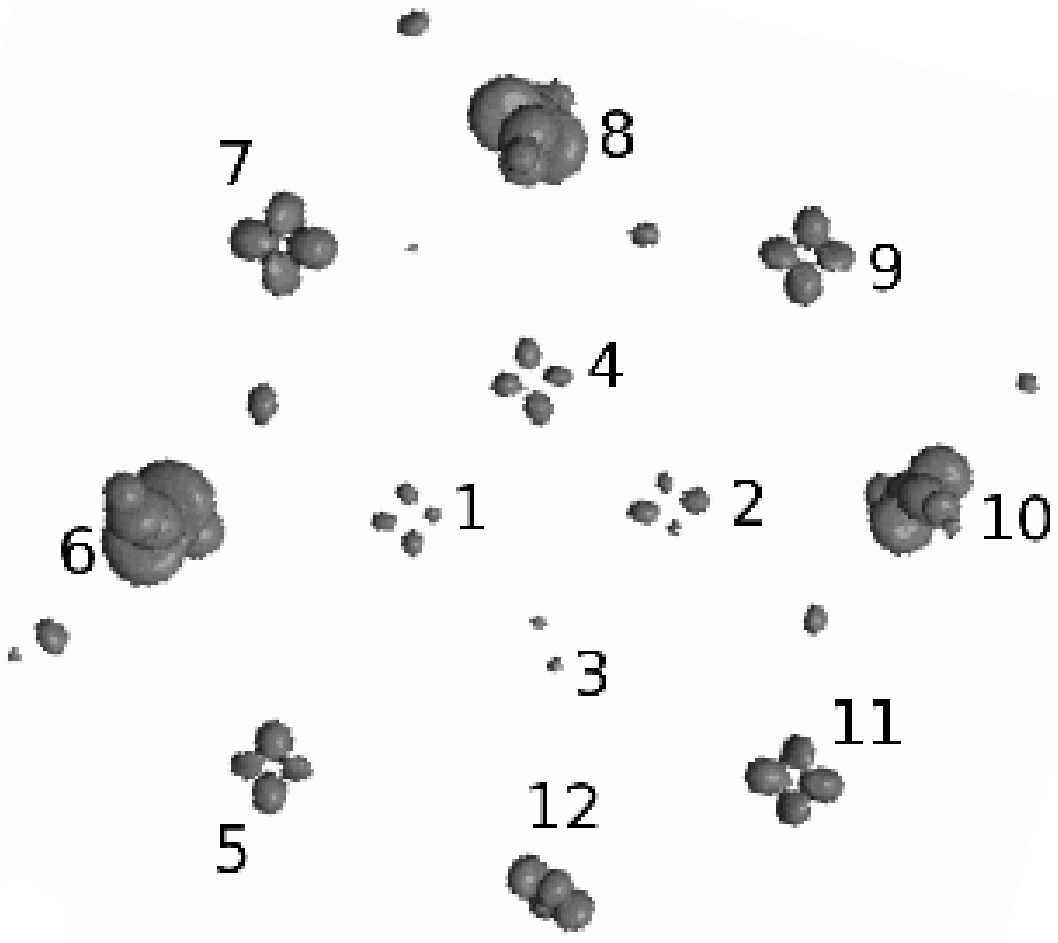}
\caption{Spatially resolved density of states (DOS) projected onto (a) the LUMO and (b) HOMO
for Geo 1:Au-(C$_3$H$_6$)$_2$, and onto (c) the LUMO and (d) HOMO for Geo 2:Au-H-no-linker.
The LUMO and HOMO originate from the majority-spin Mn $d$ orbitals. The criterion of the
isosurface shown is $2.0 \times 10^{-3}$ electrons/$a_B^3$, where $a_B$ is the Bohr radius.
The numbering of the Mn ions is based on Fig.~\ref{fig:geo}. The blobs far away from the 
core of the Mn$_{12}$ (marked by L) in (a) and (b) arise from the linker molecules.}
\label{fig:LDOS}
\end{figure}

\begin{figure}
\includegraphics[width=7.8cm, height=6.0cm]{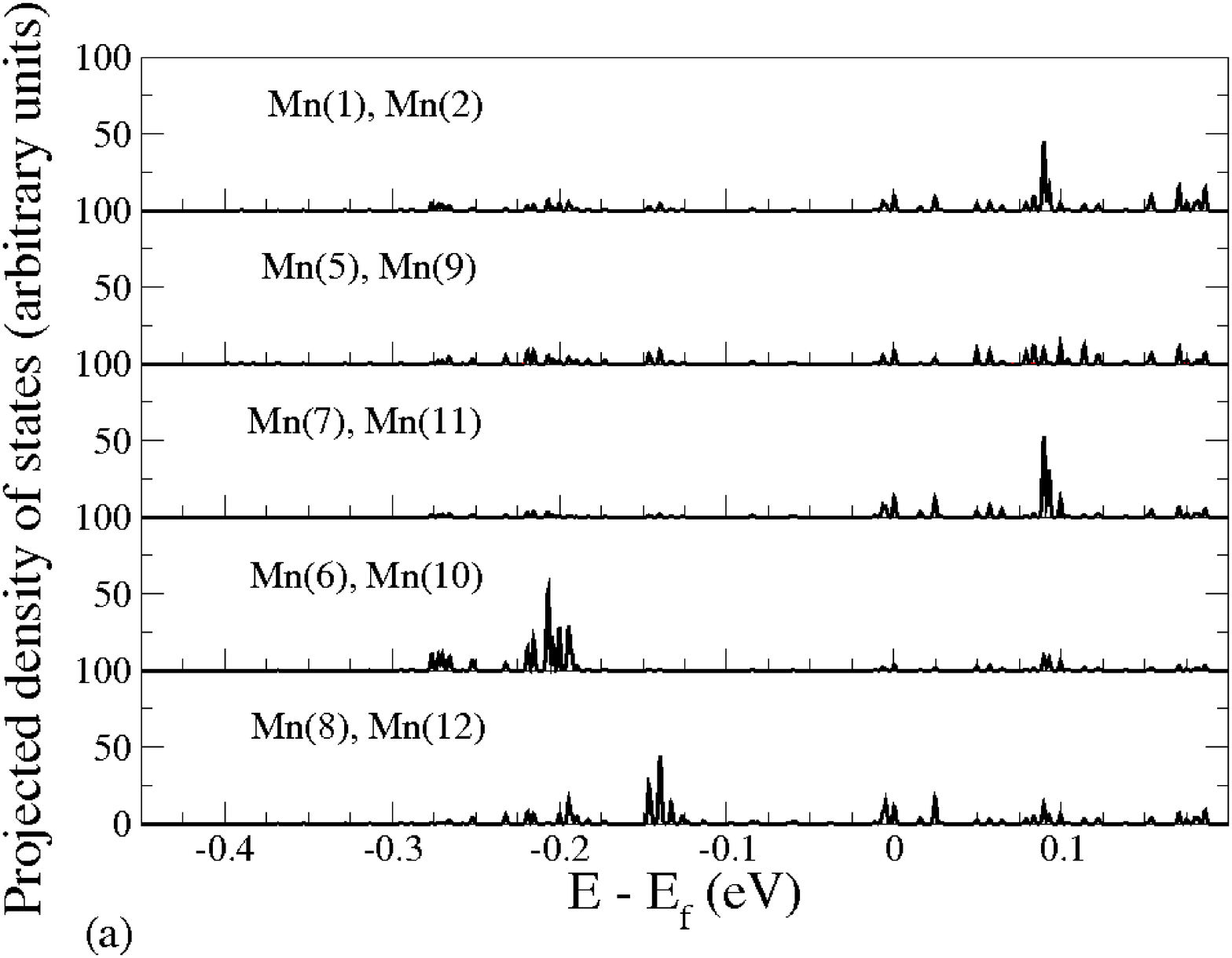}
\hspace{0.5truecm}
\includegraphics[width=7.8cm, height=6.0cm]{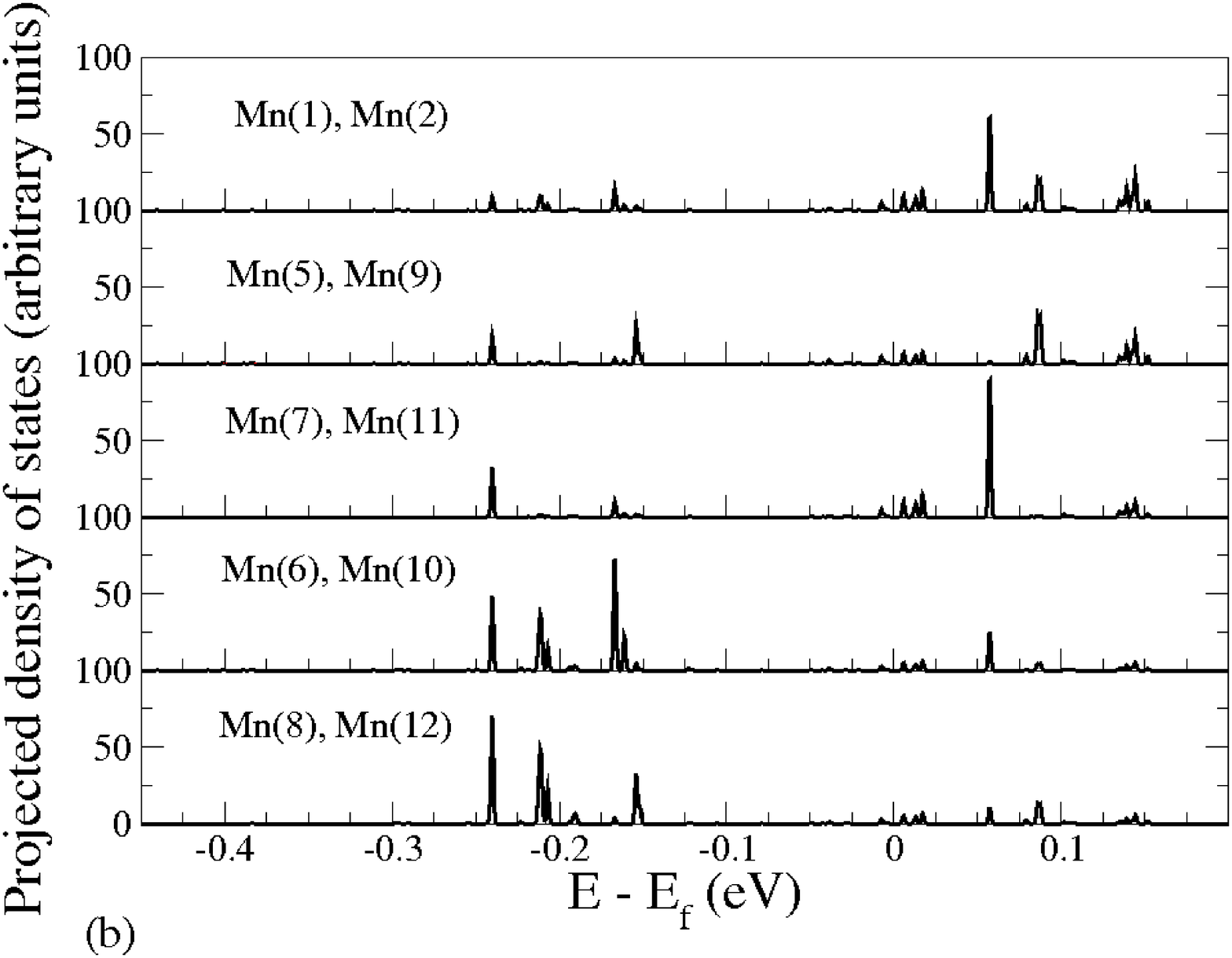}
\caption{Majority-spin density of states (DOS) projected onto Mn $d$ orbitals
for (a) Geo 1:Au-(C$_3$H$_6$)$_2$ and (b) Geo 1:Au-(AuC$_3$H$_6$)$_2$.
Refer to Fig.~\ref{fig:geo} for numbering of the Mn ions. The densities of
states for Mn(3) and Mn(4) sites are the same as those for Mn(1) and Mn(2)
sites.}
\label{fig:PDOS-geo1a}
\end{figure}

For the interface geometries with the orientation (1), Fig.~\ref{fig:geo}(a), (b), (c), and (d),
the linker molecules are closer to the Mn(5), Mn(6), Mn(9), and Mn(10) sites than the rest of the eight Mn
sites, as illustrated in Fig.~\ref{fig:geo}. Among the former four Mn sites, the Mn(5) and Mn(9) sites predominantly
contribute to the LUMO of the Mn$_{12}$ [Fig.~\ref{fig:LDOS}(a)], 
as well as to some degree to the HOMO [Figs.~\ref{fig:PDOS-geo1}(b)
and \ref{fig:LDOS}(b)]. Notice that the HOMO and LUMO arise from the majority-spin Mn $d$ orbitals.
A much greater contribution to the HOMO arises from the Mn(8) and Mn(12) sites
[Fig.~\ref{fig:LDOS}(b)], but these sites are not as close to the linker molecules as the former four sites.
Thus, for the geometries with the orientation (1), the broadening of the LUMO level is expected to be somewhat
larger than that of the HOMO level, as shown in Table~1 and Figs.~\ref{fig:PDOS-geo1}(b),~\ref{fig:TRC}(a), and
~\ref{fig:PDOS-geo1a}. For the geometries with the orientation (2), in the case of chemisorption, Fig.~\ref{fig:geo}(e),
(f), and (h), the linker molecules are in a closer proximity to the Mn(1), Mn(2), Mn(3), Mn(4), Mn(6), Mn(8), Mn(10),
and Mn(12) sites, than to the other four Mn sites, as shown in Fig.~\ref{fig:geo}. The Mn(6), Mn(8), Mn(10), and
Mn(12) sites predominantly contribute to the HOMO [Fig.~\ref{fig:LDOS}(d)] rather than the LUMO
[Fig.~\ref{fig:LDOS}(c)], and
so the broadening of the HOMO level is much larger than that of the LUMO level [Table 1 and Fig.~\ref{fig:TRC}(b)].
In Fig.~\ref{fig:TRC}(b), the transmission peak associated with the HOMO level for Geo 2:Au-S$_4$
(Geo 2:Au-H-no-linker), is very broad because it corresponds to four (two) molecular levels broadened right below
$E_f$ including the HOMO level. In the case of physisorption, Figs.~\ref{fig:geo}(g) and (i),
the Mn(6), Mn(8), Mn(10), and Mn(12) sites are much closer to the electrodes than the Mn(5), Mn(7), Mn(9),
and Mn(11) sites. The HOMO originates mainly from the former four Mn $d$ orbitals, while the LUMO comes
from the latter four Mn $d$ orbitals. Thus, for the geometries in the orientation (2), either in the case
of chemisorption or physisorption, the broadening of the HOMO level is expected to be much
larger than that of the LUMO level [Table ~1 and Fig.~\ref{fig:TRC}(b)]. 

Now we compare the level broadening for the geometries in the orientation (1) to that in
the orientation (2), for a given bonding type and fixed distance $d$. For this comparison, we examine
the following three geometries with the Au-S bonding: one geometry in the orientation (1),
Geo 1:Au-(SC$_3$H$_6$)$_2$-hollow, and two geometries in the orientation (2), Geo 2:Au-S$_4$ and
Geo 2:Au-(SC$_2$H$_4$)$_4$. The distance $d$ in the former geometry, 25.73~\AA,~is much longer than
those for the latter two geometries. Thus, we extrapolate to find the broadening of a corresponding
geometry in the orientation (2) with $d$=25.73~\AA. The Mn$_{12}$ and the linker molecules play the role
of a potential barrier in the electron transport, and so the broadening of the relevant molecular levels
decays exponentially as a function of $d$. Thus, we apply $\Gamma_{\rm LUMO}$=$C_1 \exp(-C_2 d)$,
where $C_1$ and $C_2$ are positive constants, to the two geometries in the orientation (2). Extracting
the values of $C_1$ and $C_2$, we find that the extrapolated value of $\Gamma_{\rm LUMO}$ for $d$=25.73~\AA~
is 0.0015~eV, which is less than the actual value of $\Gamma_{\rm LUMO}$ for the first geometry in the
orientation (1), 0.0020~eV. When the same logic is applied to the HOMO level, the extrapolated value of
$\Gamma_{\rm HOMO}$ for such a geometry becomes 0.0069~eV,\cite{HOMO-EXPL} which is much greater than
the actual value of $\Gamma_{\rm HOMO}$ for the first geometry, 0.00050~eV.
Consequently, for a given bonding type and fixed distance $d$, the broadening of the LUMO (HOMO) level for
geometries with the orientation (1) is more (less) pronounced than that with the orientation (2).
Combining this result with the finding that the LUMO level broadens more than the HOMO level for
the geometries with the orientation (1) and the opposite holds for those with the orientation (2),
we reach the following conclusion. Although more statistics is desirable and other bonding
types can be tested, our calculations indicate that as long as the current flows through the LUMO level,
(for a given $d$ and bonding type) geometries with the orientation (1) provide somewhat a higher current
than those with the orientation (2). However, if one can arrange the transport to occur through the
HOMO level (by application of gate voltage to the Mn$_{12}$), geometries in the orientation (2)
will give a much higher current than those in the orientation (1). One caveat is that for a short
distance $d$ the size of the Mn$_{12}$ prohibits interface geometries with the orientation (1)
from being formed.

\begin{figure}
\includegraphics[width=7.8cm, height=6.0cm]{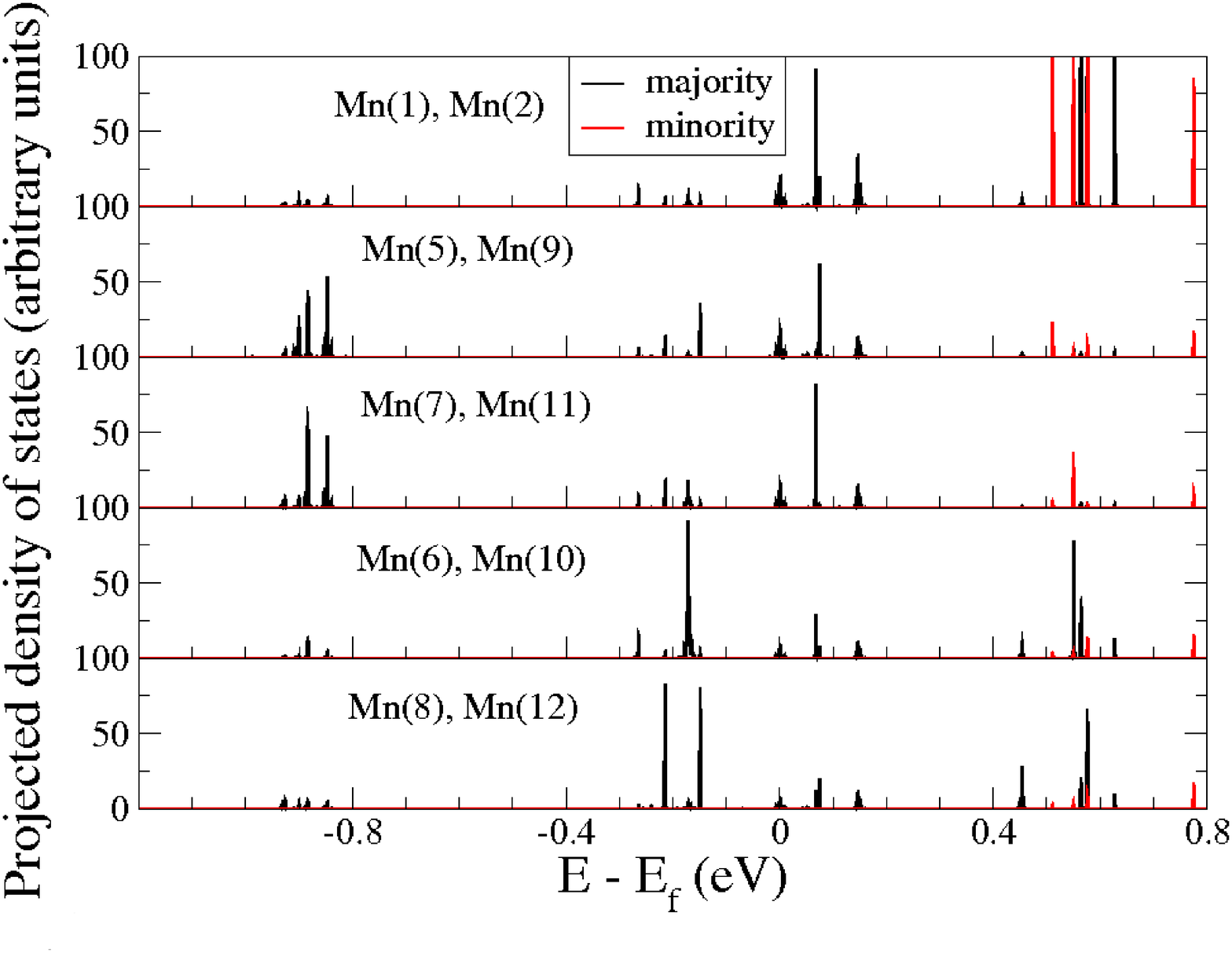}
\hspace{0.5truecm}
\includegraphics[width=7.8cm, height=6.0cm]{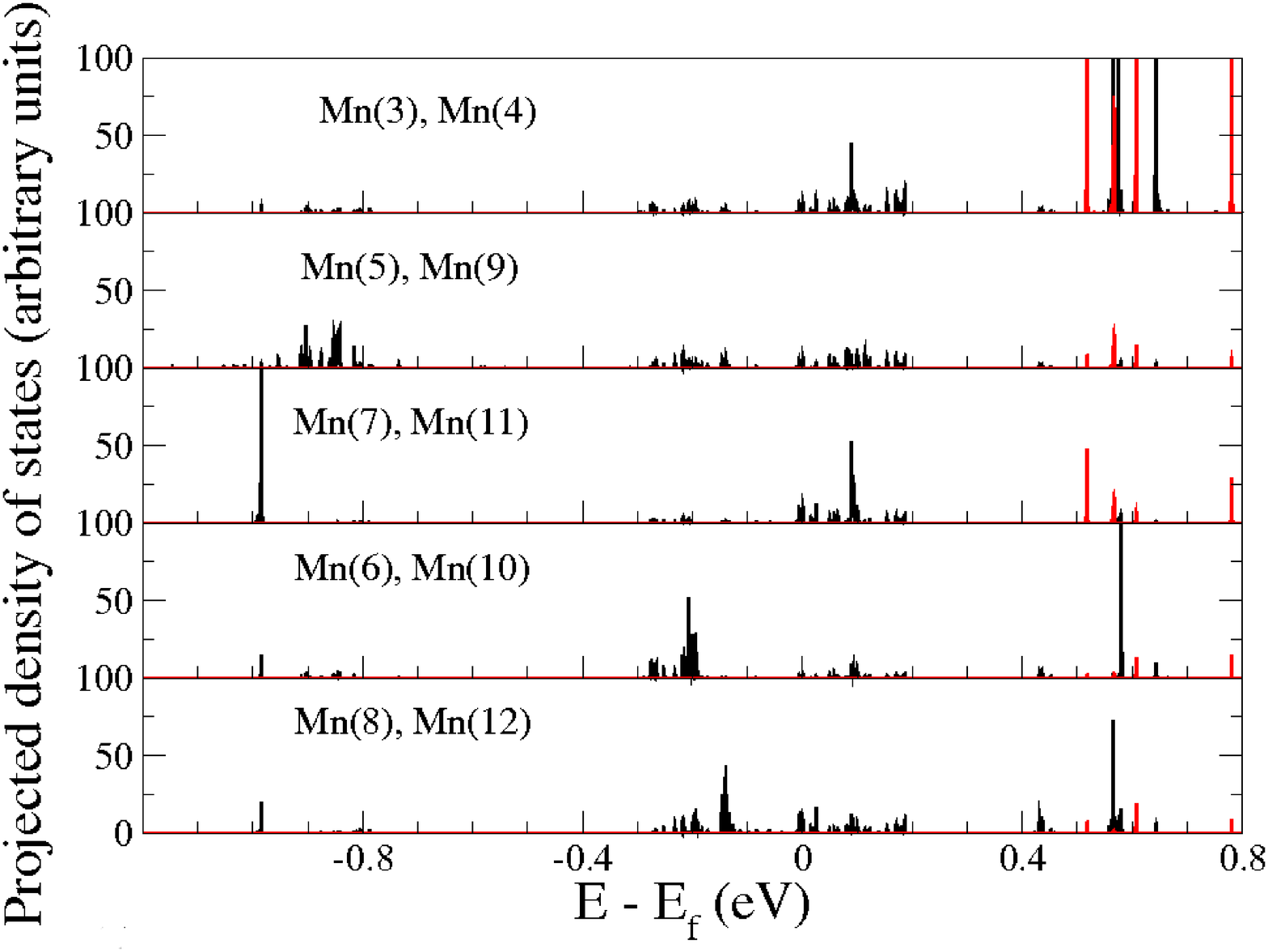}
\caption{(Color online) Spin-polarized DOS projected onto the Mn $d$ orbitals for
(a) Geo 1:Au-(SC$_3$H$_6$)$_2$-hollow and (b) Geo 1:Au-(C$_3$H$_6$)$_2$: 
majority-spin (black) and minority-spin (red).
Refer to Fig.~\ref{fig:geo} for numbering of the Mn ions. The DOS projected 
onto the Mn(1) and Mn(2) sites are similar to those onto the Mn(3) and Mn(4) sites.}
\label{fig:PDOS-SP1}
\end{figure}

\begin{figure}
\includegraphics[width=7.8cm, height=6.0cm]{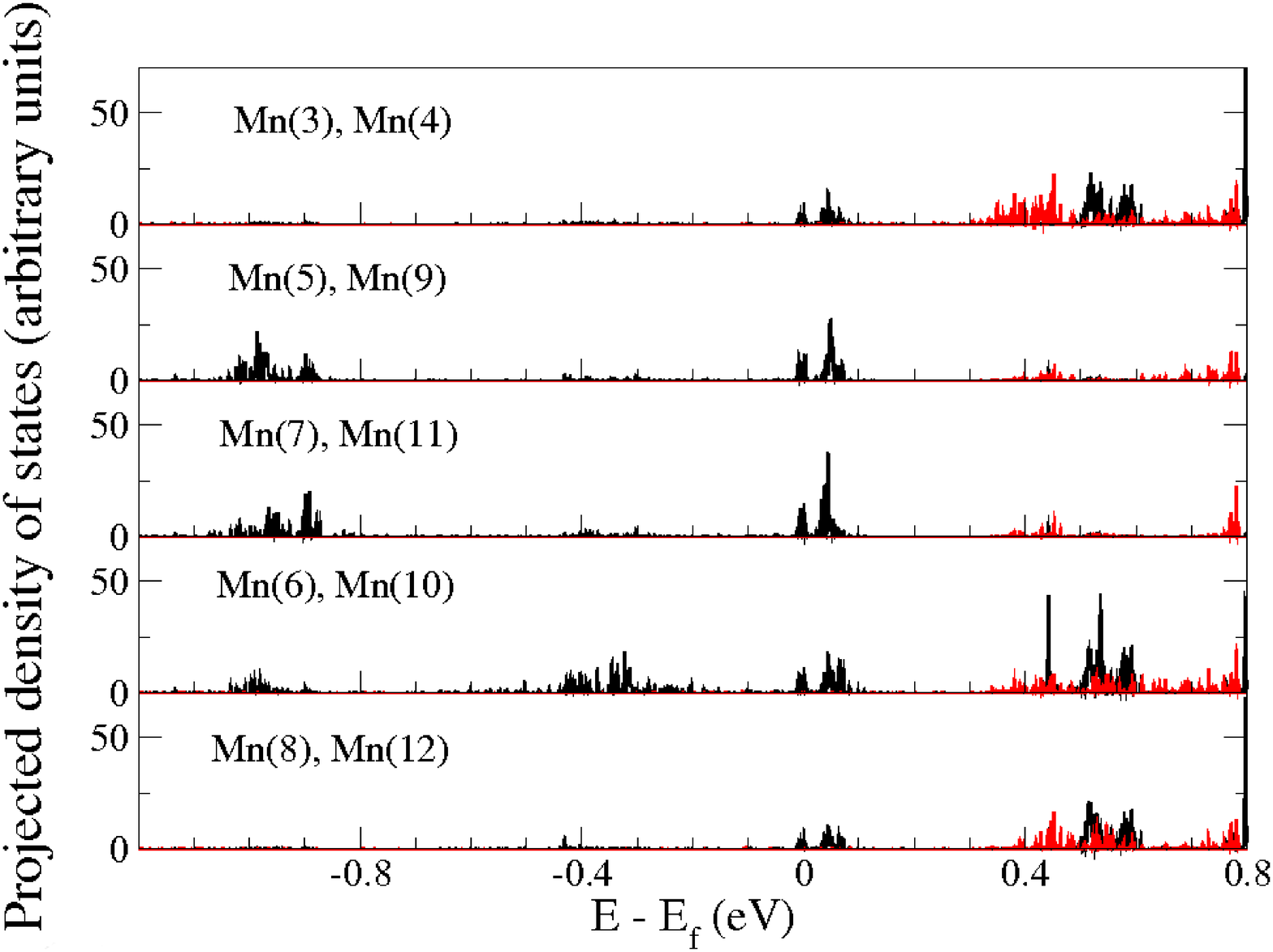}
\hspace{0.5truecm}
\includegraphics[width=7.8cm, height=6.0cm]{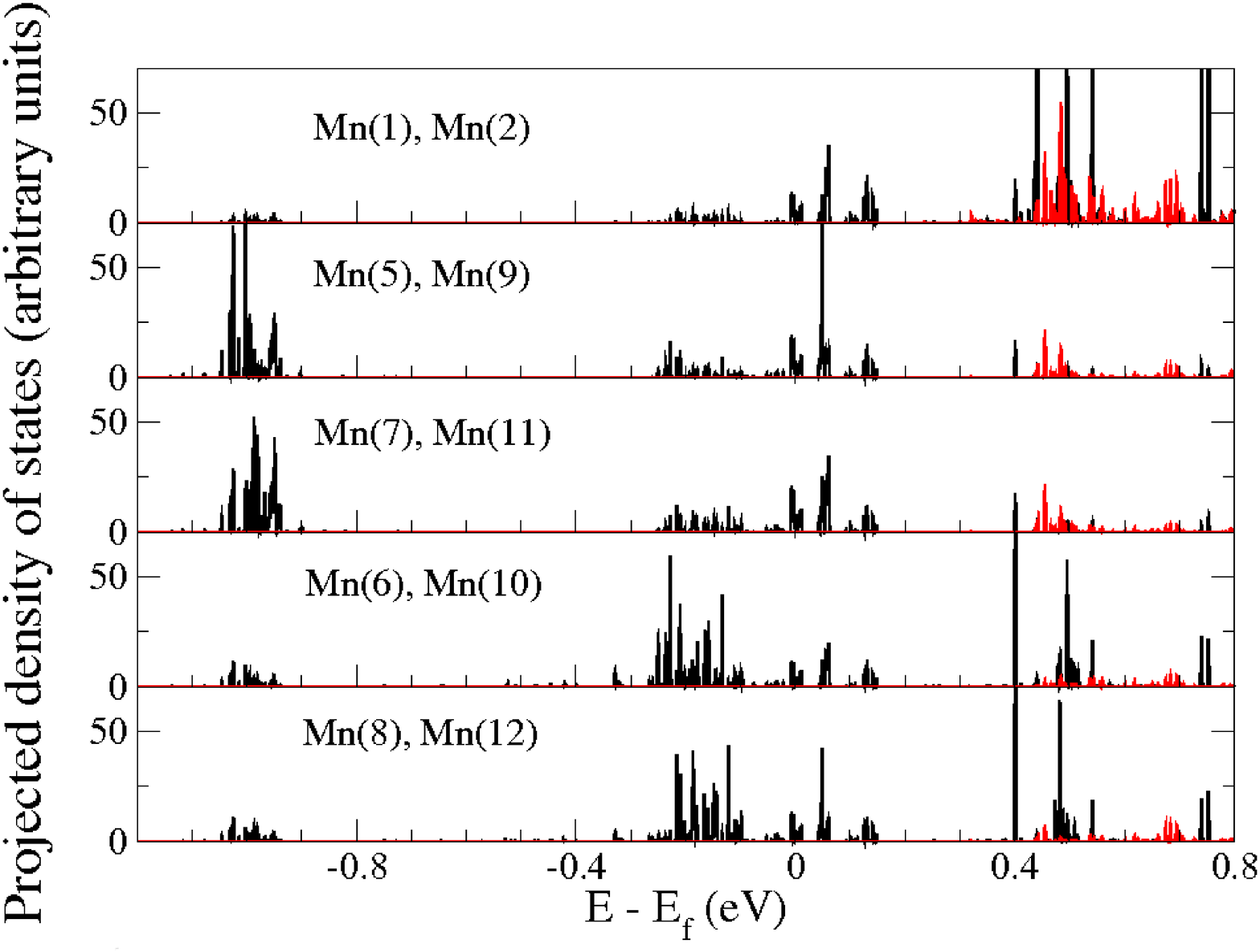}
\caption{(Color online) Spin-polarized DOS projected onto the Mn $d$ orbitals for
(a) Geo 2:Au-$S_4$ and (b) Geo 2:Au-H-no-linker: majority-spin (black) and 
minority-spin (red). Refer to Fig.~\ref{fig:geo} for numbering of the Mn ions.
The DOS projected onto the Mn(1) and Mn(2) sites are similar to those onto 
the Mn(3) and Mn(4) sites.}
\label{fig:PDOS-SP2}
\end{figure}

\begin{figure}
\includegraphics[width=7.8cm, height=6.0cm]{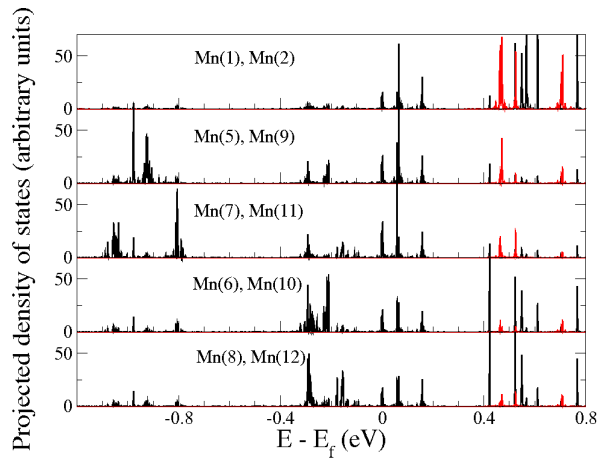}
\caption{(Color online) Spin-polarized DOS projected onto the Mn $d$ orbitals for
Geo 2:Au-(SC$_2$H$_4$)$_4$: majority-spin (black) and 
minority-spin (red). Refer to Fig.~\ref{fig:geo} for numbering of the Mn ions.
The DOS projected onto the Mn(1) and Mn(2) sites are similar to those onto 
the Mn(3) and Mn(4) sites.}
\label{fig:PDOS-SP3}
\end{figure}

\begin{figure}
\includegraphics[width=7.8cm, height=6.0cm]{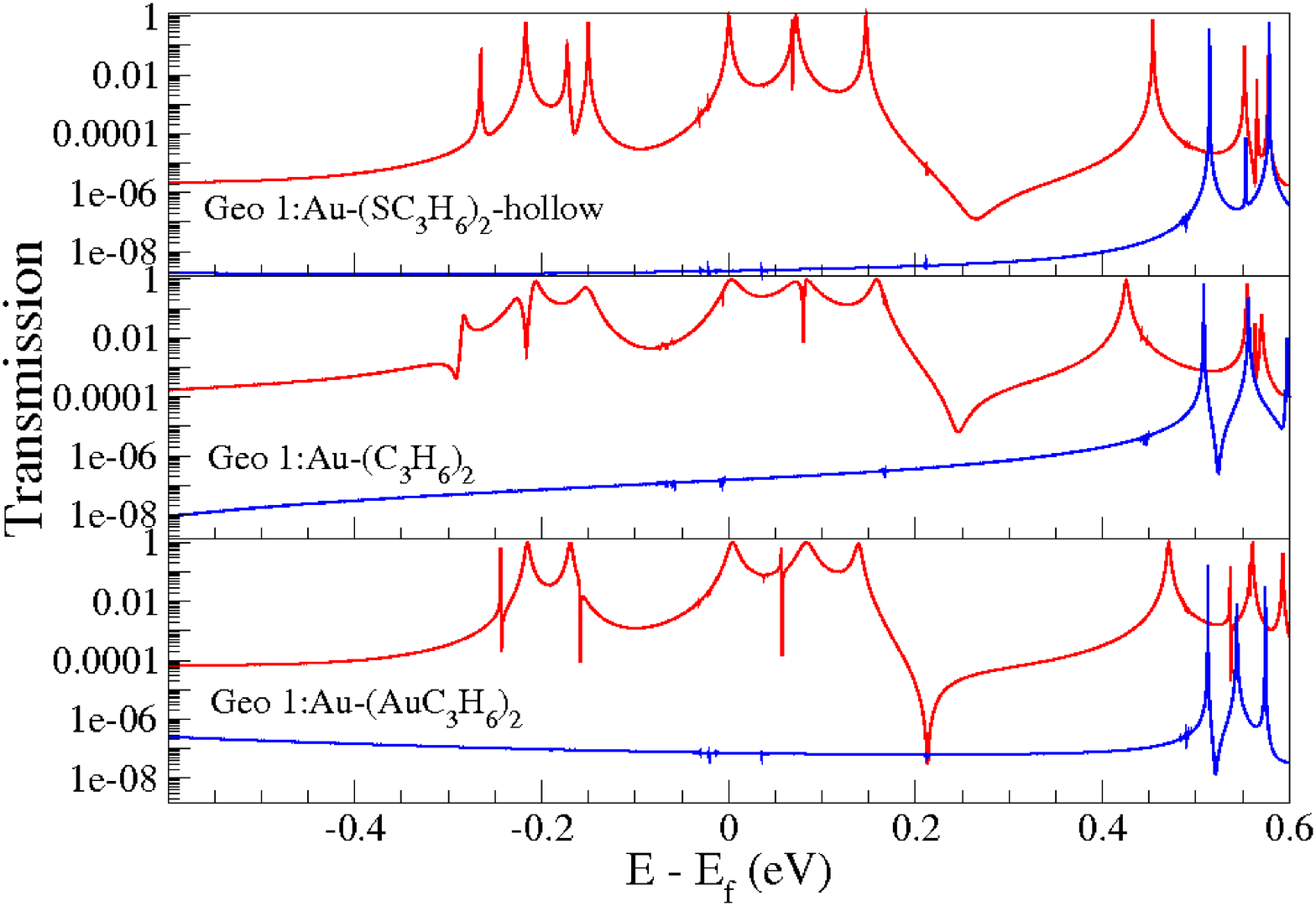}
\hspace{0.5truecm}
\includegraphics[width=7.8cm, height=6.0cm]{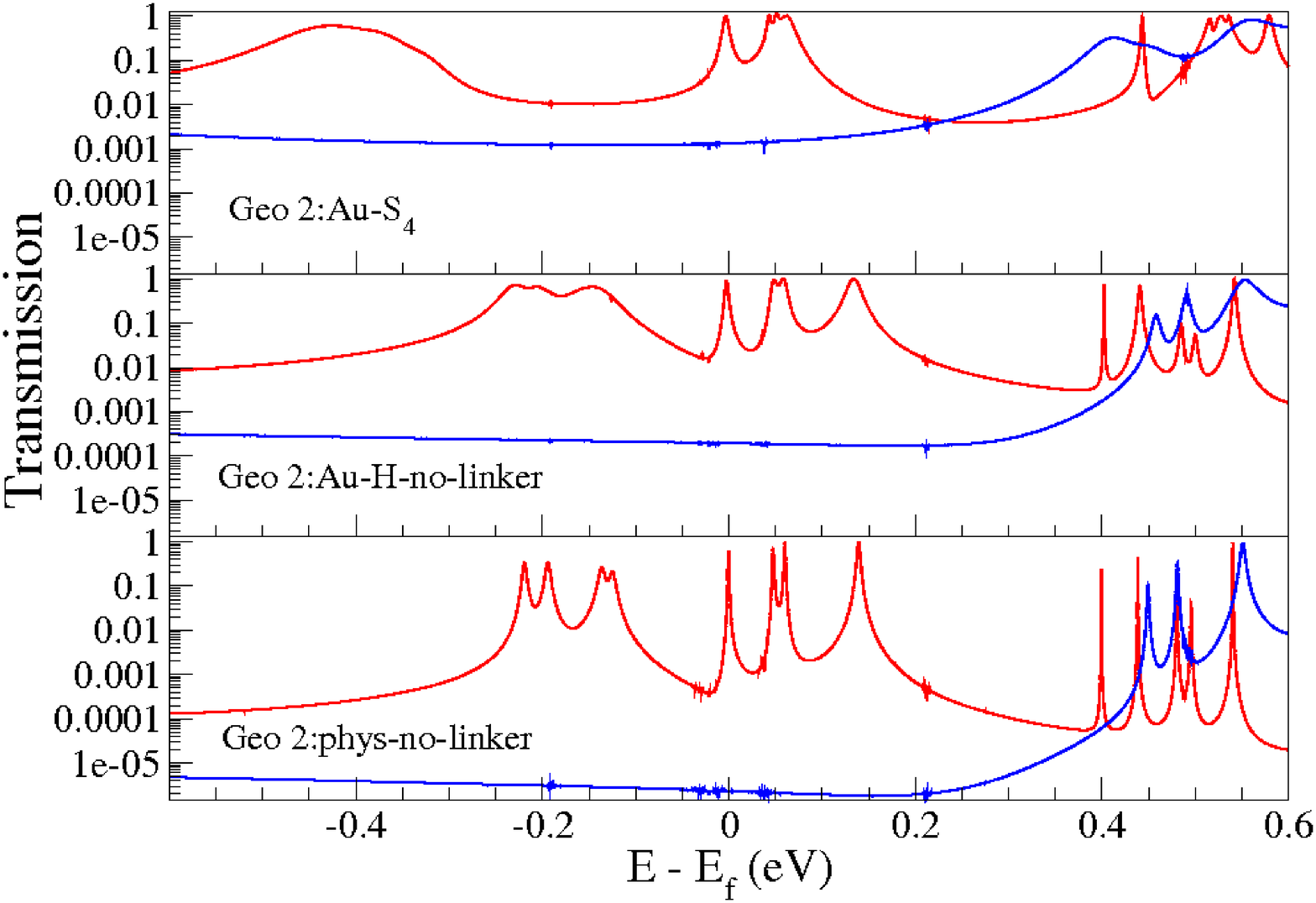}
\caption{(Color online) Spin-polarized transmission at zero bias with
a logarithmic scale in the vertical axis. Majority-spin contribution:
red, minority-spin contribution: blue. The vertical scales of the figures
on the left differ from those on the right.}
\label{fig:TRC-log}
\end{figure}

For all of the geometries considered, the effect of spin filtering remains robust. 
The spin-filtering effect occurs in the transport through the Mn$_{12}$ because
its majority-spin HOMO and LUMO levels are well separated from the minority-spin HOMO and 
LUMO levels and because the minority-spin HOMO level is located sufficiently below the 
majority-spin HOMO level (Fig. 2 in Ref.[\onlinecite{SALV09-2}], Figs.~\ref{fig:PDOS-SP1}, 
\ref{fig:PDOS-SP2}, and \ref{fig:PDOS-SP3}). This feature of the HOMO and LUMO does not
change with interface geometry. However, the degree of the spin filtering 
depends on molecular orientations, bonding types, and the distance $d$.
We first compare the spin-filtering effect for the geometries in the orientation (1) to that in the
orientation (2). Figure~\ref{fig:TRC-log} shows the spin-polarized $T(E,0)$ for six geometries
[Fig.~\ref{fig:geo}(a), (c), (d), (e), (h), and (i)]. In the last column of Table~1, the ratio of the
majority-spin to the minority-spin transmission coefficients at the energy level associated with the
LUMO level, $(T_{\uparrow}-T_{\downarrow})/T_{\downarrow}$, is provided. For the geometries in
the orientation (1), the ratio varies from 6.11$\times$$10^6$ to 5.26$\times$$10^8$, while for those
in the orientation (2), the ratio is in the range of 7.39$\times$$10^2$ to 2.86$\times$$10^5$. A larger
spin-filtering effect is obtained for the geometries in the orientation (1) than for those in
the orientation (2). This is partly because the linker molecules in the orientation (2) are in a closer
proximity to the Mn(1), Mn(2), Mn(3), and Mn(4) sites than those in the orientation (1). These four
Mn sites mainly contribute to the minority-spin density of states as shown in Figs.~\ref{fig:PDOS-SP1} 
and \ref{fig:PDOS-SP2}. Due to this proximity of the four Mn sites, the minority-spin DOS for the 
geometries in the orientation (2) is more spread than that for the geometries in the orientaion (1).
An additional reason is as follows. As discussed earlier and shown in
Figs.~\ref{fig:PDOS-SP1}, \ref{fig:PDOS-SP2}, and \ref{fig:PDOS-SP3}, 
the LUMO arises from the majority-spin Mn $d$ orbitals.
At the energy level where the ratio is obtained, the majority-spin transmission is of
the order of unity, independent of the distance $d$, because the transport is carried by the LUMO
level. However, at that energy level, the transport of minority-spin electrons is in the tunneling
regime, since there are no corresponding molecular levels of the minority-spin. Thus, the minority-spin
transmission decreases exponentially with $d$. Since the geometries in the orientation (1) have much
longer distances $d$ than those in the orientation (2), the coefficients of the minority-spin
transmission for the former geometries are much smaller than those for the latter geometries.
This renders the greatly enhanced ratio for the geometries in the orientation (1). 
Now let us briefly discuss the influence of bonding type on the spin-filtering effect.
For the orientation (1) the minority-spin DOS for the Au-C bonding is slightly broader
than that for the Au-S bonding [compare Fig.~\ref{fig:PDOS-SP1}(a) to (b)]. Thus, the
ratio for the Au-C bonding is lower than that for the Au-S bonding (Table 1) for a given
distance $d$. Similarly, for the orientation (2), the minority-spin DOS for the Au-S bonding
is more delocalized than that for the Au-H bonding [compare Fig.~\ref{fig:PDOS-SP2}(a) to (b)].
So the ratio for the Au-S bonding is lower than that for the Au-H bonding for a given distance $d$.
For a specific bonding type, the ratio increases with increasing the distance $d$ because of the
reason explained above. For example, for the Au-S bonding, the minority-spin PDOS for 
Geo 2:Au-(SC$_2$H$_4$)$_4$ is much more localized than that for Geo 2:Au-S$_4$ [compare 
Fig.~\ref{fig:PDOS-SP3} to \ref{fig:PDOS-SP2}(a)]. 
Thus, the ratio for the former is two orders of magnitude greater than that for the latter. 
Additionally, we note from Figs.~\ref{fig:TRC} and \ref{fig:TRC-log} that
this ratio, $(T_{\uparrow}-T_{\downarrow})/T_{\downarrow}$, remains in the range of 100 to 10$^7$ 
even at the energy level where the transport of both the majority spin and the minority spin is in 
the tunneling regime. The spin-filtering effect may be achieved for other SMMs with high magnetic 
moments and stable ground-state spin multiplets.

\subsection{Effect of bonding type and linker group}

We investigate the effect of binding site on transport properties for a given bonding type,
for example, the Au-S bonding in the orientation (1), Geo 1:Au-(SC$_3$H$_6$)$_2$. We consider hollow and
on-top binding sites of the S atoms to the Au surface in that geometry, without geometry relaxation. The
broadening of the LUMO level somewhat decreases from the hollow-site to the on-top geometries, while the
broadening of the HOMO level slightly increases (Table~1). This difference in the broadening is
much smaller than that induced by different bonding types. Despite this small difference, the overall
features of $T(E,0)$ for the two geometries are fairly similar to each other, as shown in Fig.~\ref{fig:TRC}(a).
Thus, the $I$-$V$ curve for the on-top site geometry is expected to be similar to that for the hollow-site geometry,
Geo 1:Au-(SC$_3$H$_6$)$_2$-hollow, shown in Fig.~\ref{fig:IV}.

We compare the transport properties of the geometries with three bonding types in the orientation (1),
where the distance $d$ is fairly similar to one another: Au-C bonding [Geo 1:Au-(C$_3$H$_6$)$_2$],
Au-S bonding [Geo 1:Au-(SC$_3$H$_6$)$_2$-hollow], and Au-Au bonding [Geo 1:Au-(AuC$_3$H$_6$)$_2$]. 
The broadening of the LUMO level for the geometry with the Au-C bonding is one order of magnitude 
larger than that with the Au-S bonding, and it is twice as large as that with the Au-Au bonding, 
as shown in Figs.~\ref{fig:TRC}(a) and ~\ref{fig:PDOS-geo1a}. Accordingly,
the current for the Au-C bonding at 50 mV is one order of magnitude higher than that for the Au-S bonding, and
is twice as high as that for the Au-Au bonding (Fig.~\ref{fig:IV}). Among all of the interface geometries
considered, for a fixed distance $d$, the geometry with the Au-C bonding reveals the largest broadening of the
LUMO level, leading to the largest current (Fig.~\ref{fig:IV}).

\begin{figure}
\includegraphics[width=7.8cm, height=6.0cm]{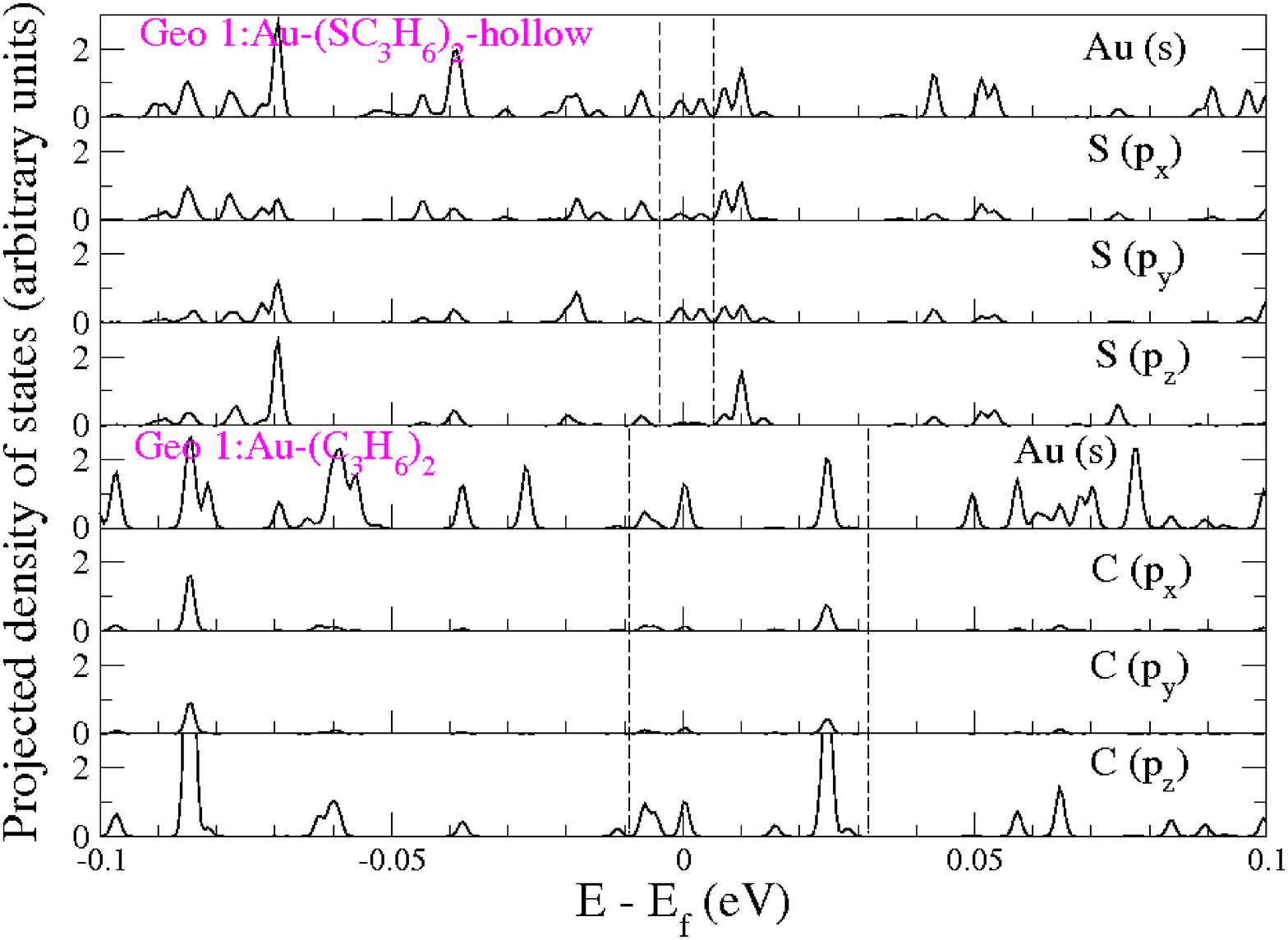}
\hspace{0.5truecm}
\includegraphics[width=7.8cm, height=6.0cm]{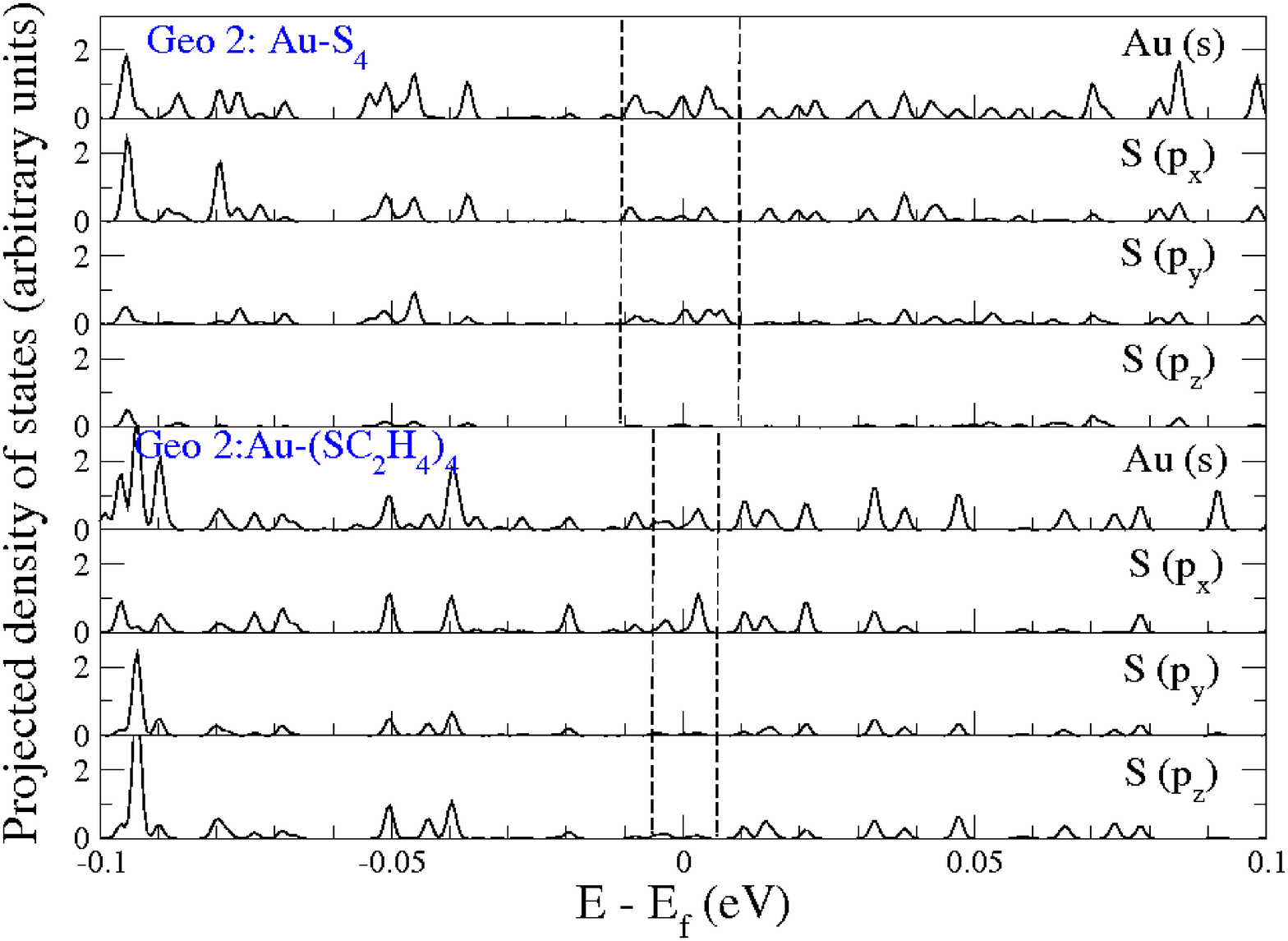}
\caption{Majority-spin DOS projected onto the $s$ orbitals of the three Au surface
atoms and $p_x$, $p_y$, $p_z$ orbitals of the S or C atom closest or bonded to the
Au atoms for (a) Geo 1:Au-(SC$_3$H$_6$)$_2$-hollow and Geo 1:Au-(C$_3$H$_6$)$_2$, and
(b) Geo 2:Au-S$_4$ and Geo 2:Au-(SC$_2$H$_4$)$_4$. In the region confined by the
dashed lines, the density peaks overlap with the LUMO.
The vertical scale in this figure differs from
those in Figs.~\ref{fig:PDOS-geo1} and \ref{fig:PDOS-geo1a}.}
\label{fig:PDOS-S}
\end{figure}

To understand the nature of the bonding between the linker molecules and the electrodes, we compute the density of
states projected onto the $p$ orbitals of the S or C atoms of the linker molecules closest to the electrodes,
and onto the $s$ orbitals of the three Au surface atoms closest or bonded to those S or C atoms. The DOS projected
onto the Au $s$ orbitals change with interface geometries due to the interactions between the linker molecules and 
the electrodes. 
For Geo 1:Au-(C$_3$H$_6$)$_2$, only the $p_z$ orbitals of the C atoms of the linker group greatly overlap with the
Au $s$ orbitals of the electrodes near $E_f$ [the bottom four panels of Fig.~\ref{fig:PDOS-S}(a)]. 
The density of the LUMO coincides with the peaks in the region bound by the dashed lines in Fig.~\ref{fig:PDOS-S}(a).
Strong overlap (in terms of the peak height and the width of the region) among the Au $s$ orbitals,
the C $p$ orbitals, and the LUMO results in a high current through the Mn$_{12}$.
For Geo 1:Au-(SC$_3$H$_6$)$_2$-hollow, the $p_x$, $p_y$, and $p_z$ orbitals of the S atoms have much weaker
overlap with the Au $s$ orbitals in the region where the LUMO appears (confined by the dashed lines) than
the Au-C bonding case [the top four panels of Fig.~\ref{fig:PDOS-S}(a)], which gives rise to a reduced transmission
probability. For Geo 2:Au-S$_4$, only the $p_x$ and $p_y$ orbitals of the S atoms bear some overlap with the Au
$s$ orbitals near $E_f$ (in the region bound by the dashed lines) [the top four panels of Fig.~\ref{fig:PDOS-S}(b)].
This overlap is weaker than that for the Au-C bonding but stronger than that for Geo 1:Au-(SC$_3$H$_6$)$_2$-hollow
or for Geo 2:Au-(SC$_2$H$_4$)$_4$ [the bottom four panels of Fig.~\ref{fig:PDOS-S}(b)]. This tendency agrees
well with the trend in the value of $\Gamma_{\rm{LUMO}}$ and the current. Additionally, our calculated spatially
resolved density of states of the LUMO [Fig.~\ref{fig:LDOS}(a)] uncovers that the geometry with the Au-C bonding 
gives rise to larger density in the linker molecules than any other geometries considered. Thus, the transmission
probability is highest for the geometry with the Au-C bonding. The geometries with the Au-S bonding do not have
as high transmission probabilities as that for the Au-C bonding or the Au-Au bonding. Consequently, the Au-C bonding
provides the highest current and the Au-S bonding gives rise to a lower current than the Au-C bonding or the Au-Au
bonding.

We discuss the transport properties of the geometries with physisorption and without linker molecules.
One of the geometries without linker molecules, Geo 2:Au-H-no-linker, has the shortest distance $d$
among the geometries considered. The broadening of the LUMO level for that geometry is 54\% of
that with the shortest Au-S bonding (Geo 2:Au-S$_4$), and it is 21\% of that with the Au-C bonding,
as shown in Table~1. The current for the geometry with the Au-H bonding at 50 mV is 44\% of that for
Geo 2:Au-S$_4$, and it is 36\% of that with the Au-C bonding (Fig.~\ref{fig:IV}).
In the case of physisorption, with the fixed distance $d$, the broadening of the LUMO level for
Geo 2:phys-no-linker is the same as that for Geo 2:phys-(CH$_3$)$_4$. This implies that the alkane chains
do not play a role in the broadening of the LUMO level. For the given distance $d$, the broadening of the
LUMO level for these two physisorbed geometries takes up only 15\% of that for the Au-S bonding
(Geo 2:Au-S$_4$), as listed in Table~1. Yet, the current for Geo 2:phys-(CH$_3$)$_4$ at 50~mV is
two orders of magnitude lower than that for Geo 2:Au-S$_4$ (Fig.~\ref{fig:IV}). Thus, in this case, the ratio
of the values of $\Gamma_{\rm LUMO}$ for the two geometries does not have the same order as the ratio of the
corresponding currents. The discrepancy between the two ratios arises from the difference
in the response of the junctions or in the changes of the transmission with bias voltage.
Notice that Fig.~\ref{fig:TRC} shows the transmission spectra for zero bias.
For chemisorbed junctions the molecular orbitals strongly overlap with the orbitals of the
electrodes. Thus, when bias voltage is applied, this overlap is enhanced especially in the regions
between adjacent transmission peaks. However, for physisorbed junctions, the transmission peaks
remain sharp even with bias voltage because of no overlap of the molecular orbitals 
with the orbitals of the electrodes. A combination of this effect with the ratio of the values of 
$\Gamma_{\rm LUMO}$ explains the ratio of the current through Geo 2:phys-(CH$_3$)$_4$ to 
that through Geo 2:Au-S$_4$. 
Our overall comparison of the transport behavior among different bonding types summarizes that the current
through the Mn$_{12}$ decreases in the following order, for a given distance $d$: Au-C, Au-Au, Au-S, Au-H,
and physisorption.


\section{Conclusion}

We simulated a single-molecule junction based on the SMM Mn$_{12}$ connected to Au(111) electrodes via
five different bonding mechanisms and two molecular orientations, and calculated their transport properties,
using the non-equilibrium Green's function method and the spin-polarized DFT. Geometry relaxation of the
junction renders a small decrease in the electric current at low bias voltage.
If the electron transport is carried by the LUMO (HOMO) level of the Mn$_{12}$, geometries in the orientation
(1) provide a higher (lower) current that those in the orientation (2).
This is due to the structure of the Mn$_{12}$. Our two-terminal transport calculations reveal that
the LUMO level is relevant to the transport through the Mn$_{12}$. In all of the interface geometries
considered, the spin-filtering effect discussed in Ref.~[\onlinecite{SALV09-1}] remains robust, and
it may occur to some other SMMs where the ground-state spin is large and the ground-state spin
multiplet is reasonably well separated in energy from the low-lying excited spin multiplets.
Some experimental studies \cite{BURG07,VOSS08,VOSS07,OTER09} show that upon deposition of Mn$_{12}$ molecules
on an Au surface, the valence states of all of the Mn ions are preserved, while some other experimental works
\cite{ZOBB05,DELP06,MANN09} reveal that such a deposition induces changes in the valence states of some of the
Mn ions. In cases where such changes do not occur \cite{BURG07,VOSS08,VOSS07,OTER09}, the
spin-filtering effect is expected. We also find that the current through the Mn$_{12}$ depends on a bonding
type and that it decreases in the following order for a fixed $d$: Au-C, Au-Au, Au-S, Au-H, and physisorption.
This is because the overlap among the LUMO, the S or C $p$ orbitals from the linker molecules, and the 
$s$ orbitals of the Au surface atoms, decreases in that order. 
The degeneracy in the magnetic levels of the Mn$_{12}$ for a given spin $S$
is lifted by spin-orbit coupling. Then inelastic transport through low-energy spin excitations must be
included for quantitative comparison to experimental data. However, the general trend in the transport
as a function of bonding type and interface geometry found in this study may still hold.

\begin{acknowledgments}
K.P. is supported by NSF DMR-0804665 and the Jeffress Memorial Trust Funds.
V.M.G.S. thanks the Spanish Ministerio de Ciencia e Innovaci\'on for 
a Juan de la Cierva fellowship and the Marie Curie European ITNs FUNMOLS 
and NANOCTM for funding.
J.F. is supported by MEC FIS2006-12117.
Computational support was provided by the SGI Altix Linux Supercluster
(Cobalt) and Intel 64 Cluster (Abe) at the National Center for Supercomputing
Applications under DMR060011 and by Virginia Tech Linux clusters and Advanced
Research Computing.
\end{acknowledgments}

\clearpage

\begin{table}
\begin{center}
\caption{Average broadening of the HOMO and LUMO levels ($\Gamma_{\rm{HOMO}}$,
$\Gamma_{\rm{LUMO}}$) of the Mn$_{12}$ for the nine interface geometries.
All geometries are optimized, unless stated otherwise. The broadening is calculated
from zero-bias transmission spectra. The value of $\Gamma_{\rm{HOMO}}$ for
Geo 2:Au-S$_4$ (Geo 2:Au-H-no-linker) is marked by $\dag$ because it represents
the full width at half maximum of the single broad transmission peak associated with
the four (two) molecular levels right below $E_f$ including the HOMO level, as shown in Fig.~\ref{fig:TRC}(b).
The ratio of the majority-spin to the minority-spin transmission, $(T_{\uparrow}-T_{\downarrow})/T_{\downarrow}$,
is computed at the energy level corresponding to the LUMO level.}
\label{table:1}
\begin{ruledtabular}
\begin{tabular}{c|c|c|c|c|c|c}
interface geometry  & bonding type & $d$ (\AA)~ & Fig.~1 & $\Gamma_{\rm{LUMO}}$ &
$\Gamma_{\rm{HOMO}}$ & $(T_{\uparrow}-T_{\downarrow})/T_{\downarrow}$ \\
(orientation, linker group) & binding site & & & (eV) & (eV) & \\ \hline
Geo 1:Au-(SC$_3$H$_6$)$_2$-hollow & Au-S, hollow & 25.73 & N/A & 0.0028 & 2.6$\times$$10^{-5}$
& 2.57$\times$$10^8$ \\
initial geometry & & & & & & \\ \hline
Geo 1:Au-(SC$_3$H$_6$)$_2$-ontop & Au-S, on-top & 25.73 & (b) & 0.0020 & 3.4$\times$$10^{-5}$
& 5.26$\times$$10^8$ \\
 initial geometry & & & & &  & \\ \hline
Geo 1:Au-(SC$_3$H$_6$)$_2$-hollow & Au-S, hollow & 25.73 & (a) & 0.0020 & 0.00050 & 4.37$\times$$10^8$ \\ \hline
Geo 1:Au-(C$_3$H$_6$)$_2$ & Au-C, on-top & 23.33 & (c) & 0.020 & 0.020 & 6.11$\times$$10^6$ \\ \hline
Geo 1:Au-(AuC$_3$H$_6$)$_2$ & Au-Au, hollow & 25.73 & (d) & 0.0090 & 0.0028 & 1.40$\times$$10^7$ \\ \hline
Geo 2:Au-S$_4$ & Au-S, hollow & 14.48  & (e) & 0.0078  & 0.11$^{\dag}$ & 7.39$\times$$10^2$  \\ \hline
Geo 2:Au-(SC$_2$H$_4$)$_4$ & Au-S, hollow & 19.01 & (f) & 0.0040 & 0.016 & 2.86$\times$$10^5$  \\ \hline
Geo 2:phys-(CH$_3$)$_4$ & physisorption & 14.56 & (g) & 0.0010 & 0.0032  & 2.32$\times$$10^5$ \\ \hline
Geo 2:Au-H-no-linker & Au-H, on-top & 12.69 & (h) & 0.0042 & 0.054$^{\dag}$ & 4.63$\times$$10^3$ \\ \hline
Geo 2:phys-no-linker & physisorption & 14.48 & (i) & 0.0011 & 0.0092 & 2.74$\times$$10^5$
\end{tabular}
\end{ruledtabular}
\end{center}
\end{table}


\begin{thebibliography}{10}
\bibitem{HOMO-EXPL}
For Geo 2:Au-S$_4$, the broadening of the HOMO level of the Mn$_{12}$ is buried under the single
broad transmission peak (width of 0.11~eV) associated with four molecular orbitals right below
the Fermi level including the HOMO level. Thus, in order to compute $C_1$ and $C_2$ for the HOMO 
level, we assume that the value of $\Gamma_{\rm HOMO}$ is $0.11/4$~eV.
\end{thebibliography}
\end{document}